# Bridging the gap between micro-economics and micro-mobility: a two-dimensional risk-based microscopic model of pedestrians' and bicyclists' operational behaviors

Mohaiminul Haque[a], Samer Hamdar[a] and Alireza Talebpour[b]

[a]*The George Washington University, mohaiminul@gwu.edu*
[a]*The George Washington University, hamdar@gwu.edu*
[b]*University of Illinois at Urbana Champaign, ataleb@illinois.edu*

**Abstract**

Due to the inherent safety concerns associated with traffic movement in unconstrained two-dimensional settings, it is important that pedestrians' and other modes' movements such as bicyclists are modeled as a risk-taking stochastic dynamic process that may lead to errors and thus contacts and collisions. Among the existing models that may capture risk-taking behaviors are: 1) the social force models (through the interplay of the repulsion and the attraction force parameters); 2) and the discrete-choice models (through the rationality or the bounded rationality paradigm while weighing different alternatives). Given that the social force models may not readily capture the contact/collision dynamics through the Newtonian force framework, decision-making theories are hypothesized as a feasible approach to formulate a new model that can account for cognitive and behavioral dimensions such as uncertainty and risk. However, instead of relying on the bounded rationality theory, in this paper, a generalized Prospect Theory based microsimulation model is proposed. The model relies on the micro-economics Prospect Theory paradigm where pedestrians or bicyclists (i.e., micro-mobility users) evaluate their speed and directional alternatives while considering the possibility of colliding with other obstacles/users. A numerical analysis on the main model parameters is presented. The model is then calibrated and validated using two real-world data sets with trajectories recorded in naturalistic settings. With the calibrated parameters studied, simulation exercises and sensitivity analysis are conducted to recreate bottlenecks and lane formations in different conditions. The findings show that the proposed model's parameters reflect the risk-taking tendencies of different roadway users in mixed right-of-way's environments while showing realistic microscopic and macroscopic traffic flow characteristics.

*Keywords:* Bicycles, Pedestrians, Prospect Theory, Risk, Safety, Traffic Simulation, Uncertainty.

# 1 Introduction

Urban planners and governments are faced with the dilemma of providing high-quality mobility services to a growing population, while at the same time minimizing energy consumption, reducing harmful environmental impacts, and cultivating a lively and safe urban quality of life. As a mean of meeting these challenges, the bicycle has resurfaced as a valuable transportation mode. Policy revisions and infrastructure amendments aimed at increasing bicycle use have led to a significant increase in the bicycle modal share in many urban areas. As a result, the traffic composition is becoming increasingly heterogeneous, with bicyclists, motor vehicles' drivers/passengers, pedestrians, and other road user groups sharing the road space. Accommodating for the needs of many road user groups with widely varying characteristics while protecting vulnerable road users, including bicyclists and pedestrians, have become a key challenge in urban transportation. For example, the physical and kinematics/dynamics (e.g., size, steering flexibility …etc.) differentiate bicyclists from other motor vehicles. Because bicycles are narrow, they can utilize lateral space within a traffic lane or bicycle lane to a greater degree than other motor vehicles. This lateral flexibility requires a deeper understanding of the influence of leading vehicles/pedestrians on the movement of bicyclists. Empirical data indicate that situations seldom occur where the movement of a bicyclist is limited by the speed and behavior of leading road users over a longer time period *(1, 2)*. Not only do bicyclists have a greater degree of lateral flexibility within a lane, but they can also switch easily between different types of available infrastructure (e.g., roadway, bicycle lane, or sidewalk). This can lead to bicyclists riding in discordance with the traffic laws (e.g., against the traffic flow, on sidewalks, in pedestrian zones, and so forth). Research indicates that in Germany, 8–20% of bicyclists ride in the wrong direction, depending on the infrastructure, and 2–15% ride on the sidewalk *(3, 4)*. Alrutz et al. *(3)* found that many bicyclists do not stop at red lights, especially at intersections where a major road crosses a minor road. In Australia, it was found that 6.9% of commuting cyclists run red lights *(5)*. The actual portion of bicyclists that break traffic rules likely depends strongly on the bicycle and driving infrastructure and the mobility culture. These differences – including the difference in risk-taking tendencies of bicyclists – must be considered and reflected in the modeling of mixed traffic, to realistically depict the resulting dynamics through traffic simulations. In other words, the objective of this paper is to develop a multidimensional microscopic traffic model for pedestrians, and bicycles that can capture the behavior of pedestrians and bicyclists in a shared space. This model allows collision formation while incorporating perception and risk-taking parameters in the decision-making process. This model will assist engineers and planners in creating safer and smarter urban transportation network. Towards realizing such objective, a micro-economics behavior-based formulation is offered and tested in terms of its feasibility. The formulation is then translated into a mathematical model that is computationally inexpensive and incorporates a cognitive decision-making process that is based on Prospect Theory. Though the proposed model is inherently capable of reproducing collisions and can be adapted to capture multimodal mixed environment scenarios with motorized and non-motorized vehicles, the focus of this paper is on the two-dimensionality of the interactions of two micro-mobility modes (pedestrians and bicycles) in a mixed non-marked environment. This focus poses multiple challenges. *i)* The traffic trajectories are not homogeneous and can differ dramatically depending on the pedestrians'/bicyclists' risk-taking attitudes. *ii)* Moreover, modeling such trajectories is computationally "expensive" with multiple behavioral and physiological/dynamics characteristics to be considered. *iii)* Finally, collecting trajectory data at the moments that lead up to a pedestrian-bicyclist interaction/collision is challenging in nature. Such trajectory data collection is needed for calibration and validation purposes. At this stage, collision reproducibility and safety analysis are out of the scope of this formulation/calibration stage of the research. In order to address the three above challenges, the following solution approach is suggested: 1) we propose a generalized traffic flow model that is based on micro-economics where utilities and outcome probabilities (i.e., collision probabilities) feed into risk-taking tendencies to be considered by decision makers. 2) We implement the model in a mixed space environment while adopting heuristics and computational methods to ensure simulation efficiency. 3) Finally, we will calibrate the model using actual trajectories of pedestrians and bicyclists mined from the Netherlands *(6)* and collected from the National Mall area in Washington DC, USA. Some e-scooterists were observed in the recorded trajectories but the associated data-points were not sufficient to conduct a thorough calibration exercise for this type of roadway users.

The first major contribution of this research is capturing pedestrian/bicyclist behavior while incorporating roadway users' risk-taking attitude in the model equations. The model formulated in this paper does not exogenously impose safety constraints to prevent accidents. Models used in practice typically preclude accidents, contrary to real-life situations. It should be mentioned that Hamdar et al., 2021 *(39)* presented a related formulation but the formulation had a very simplistic collision probability assumption and was not used to reproduce real-world trajectory data in different conditions. Moreover, the formulation was only for pedestrians. The second contribution of this research is calibrating the proposed model's parameters with an existing and a newly collected trajectory data sets associated with two different naturalistic settings. The calibration exercise shows the reflection of the models' parameters of the risk-taking tendencies of pedestrians and bicyclists while dealing with different contact/collision types. The values of these parameters may be seen as a new family of endogenous surrogate safety measures that are inherent to the proposed model's structure while still producing the expected macroscopic and microscopic flow properties.

Given the aforementioned contributions, the steps performed in this research and shown in this paper are:

a- Proposing a Prospect Theory-based formulation for pedestrians/bicycles and testing the model in terms of flow-density-speed macroscopic properties.
b- Utilizing a suitable calibration methodology and making sure that the suggested model reproduces rear-world trajectories (with focus on pedestrians and bicycles given the available trajectory data sets) especially during different traffic conditions.
c- Investigating the macroscopic relationship of the presented calibrated model with different values of collision weight.

Towards realizing these steps, a literature review of pedestrian and bicyclist models (i.e., same modeling framework for these two micro-mobility modes) is presented in the following section. Section 3 presents the micro-economic behavior model of pedestrians/bicyclists in a shared space with the ability to reproduce collisions. After calibrating the model in Section 4, the fifth section includes the simulation results and the corresponding data analysis before concluding with a summary remarks and future research needs.

## 2 Literature Review

With the growth of mixed land use development across the world and the rise of popularity of different motorized and non-motorized shared transport services like Bicycle/E-bicycle share, E-scooter share etc., the diversity of road users is increasing *(7)*. When it comes to micro-mobility services, these diverse road users with unique features of their own are often sharing the same roadway and sidewalk with the more traditional road users such as car/bus drivers on roadways and pedestrians on sidewalks. This mix without pre-defined regulations and rights of ways are raising significant operational and safety concerns *(8)*. To mitigate these concerns and to estimate the effect of policies or traffic operation plans regarding such new category of road users (bicyclists and e-scooterists) in the presence of other vulnerable users such as pedestrians, decision makers need to understand the dynamics of mixed (i.e., with multiple modes) micro mobility two-dimensional traffic sharing the same space (i.e., with no clear right-of-way – ROW). This section will first present common microscopic operational pedestrian (i.e., walking) models and bicycle models. It will also offer some review on existing common frameworks for modeling pedestrian and bicycle models given the nature of the model proposed in this manuscript. This section is not meant to offer an exhaustive review of literature on all existing pedestrian, bicycle, and e-scooter operational microscopic models but a robust review to highlight the significance of the research offered in this manuscript.

In recent years, the study of pedestrian traffic and bicycle traffic has attracted wide attention in the fields of physical science and engineering. Pedestrian walking behavior and operational movement is common in daily life and complicated aggregate pedestrian dynamics such as the lane formation in bidirectional flow, and turbulent movement in dense crowds can be observed. Many empirical studies focus on a fundamental diagram of the relationship between speed, density, and flow *(9-12)*; in addition, experiments have investigated some self-organization phenomena that are based on one-to-one or one-to-many interactions (*13*). Many models have been proposed or revised to simulate the observed pedestrian flow characteristics, such as the continuum model *(14)*, the lattice gas model (*15*), the force field model (*16*), the social force model (*17*), the heuristics-based model (*18*), and the utility optimization-based model (*44*). Most of these models did not explicitly focus on cognitive decision-making processes such as risk-taking and sensitivity to stimuli.

In addition to the pedestrian walking behavior models, models have been proposed to represent bicycle traffic flow microscopically (i.e., at the operational level). Among the first microscopic models for bicyclist operations were those that used modelling paradigms developed for cars. One example are the Cellular Automata models, whose parameters have been adjusted to capture the smaller size and lower speeds of bicycles compared to cars (*19-21*). However, the rules that determine movements between cells have not been modified to represent cycling behavior and interactions. In a lane-based, one-dimensional flow, a bicycle-following model has been derived and calibrated using empirical bicyclist data *(22)* while Zhao and Zhang (2017) *(23)* showed that a unified following model can be applied to cars, pedestrians and cyclists and produce realistic macroscopic characteristics given proper scaling of the parameters. Despite this finding, single file bicycle flow is generally not representative of bicyclists' movements on cycling infrastructure. Due to the lack of lanes, identifying bicyclist overtaking maneuvers is not trivial, but requires specific thresholds to be differentiated from a following movement (*24*). Apart from following and overtaking, there are more situations that a microscopic bicyclist flow model should cover. These relate to the individual bicycle movement and the encounters with traffic participants from other directions. In (*25*), the acceleration process of bicyclists using naturalistic data is modelled; (*26-27*) developed social force models that consider repulsive forces from obstacles and other traffic users, as well as an ellipse representation of the bicycle. In (*45*) a game theory-based bicycle flow model is presented which takes the bicycles' kinematic properties into consideration. In *(28)* a two-layer framework is proposed to model operational cycling behavior using discrete choice theory and is demonstrated on bicycle interactions while forming a queue upstream of a traffic signal. As in the case for pedestrian flow models, the aforementioned bicyclists' models lack cognitive representation and do not consider collision formation as an outcome of a balance between reaching a given destination point while avoiding collisions/contacts for safety reasons.

Though simulation have long been used as a useful tool to understand traffic dynamics of a single type of road users (vehicular flow models, pedestrian flow models etc.), the effort to model and simulate mixed traffic is comparatively very limited. To the authors' best knowledge, several researchers have proposed simulation frameworks for mixed traffic of two or more models till date. We can broadly classify them into two categories. First, researchers proposed simulation of mixed traffic with different models for different traffic agents (cars, bicycles, pedestrians etc.). For example, *(29)* presented an agent-based simulation framework for cars, trams and pedestrians. For cars and trams' agents, the authors used a generalized force model; a discrete choice model was adopted to simulate the movement of pedestrians' agents. *(29-31)* used the Social Force Model (SFM) with different car following models to model pedestrian-vehicle interactions. *(32)* proposed an integrated simulation model by combining a 1D car-following model and a 2D floor field pedestrian model for modeling pedestrian-vehicle interactions and integrating them in a comprehensive agent-based framework. *(33)* used a large-scale traffic simulation framework based on the queueing model of *(34)* to integrate a multi-modal simulation module to the existing framework of MATSim. A force-based 2D simulation module was used for the non-vehicular trips. *(35)* introduced pedestrian and bicycle agent models into the open-source traffic simulation SUMO. For pedestrian and bicycle agents, the authors used relatively simple behavior rules.

A second category of models was developed by researchers by changing an existing modeling framework and parameters to investigate the possibility of simulating the traffic dynamics of a different transportation mode. For example, *(31)* also developed a 2D car behavior model based on SFM. The proposed approach uses a mechanistic module to obtain reasonable turning trajectories; a proportional-integral-derivative (PID) controller is integrated to control the simulated vehicles. *(36)* attempted to model mixed traffic flow at signalized intersections with an extended SFM model. *(37)* presented a bicycle simulation model by reparametrizing the Intelligent Driver car-following model (IDM). The authors hypothesized that there is no qualitative difference between vehicular and bicycle traffic flow dynamics and found similar quality metrics for both models.

From this literature review, a main conclusion may be stated: the existing modeling efforts of mixed traffic in shared spaces are somehow successful in generating observed mobility patterns and flow dynamics. However, none of the existing modeling efforts is based on decision-making theories with parameters associated with defined cognitive dimensions (especially risk). Moreover, these models have an inherent inability to reproduce collision in mixed environments (mainly when motorized vehicles are involved) since they are based on collision-free models that were extended for alternative applications. The micro-economic based model presented in this paper has the ability to reproduce collisions and can be calibrated to reproduce real-world trajectory patterns. It may be applied for a car-following formulation but with significant extensions to account for directionality. This comprehensive modeling approach makes the proposed model suitable for behavioral comparisons associated with the risk-taking tendencies of different roadway users (e.g., perception regarding collisions between two cars by a driver versus perception regarding collision between a bicycle and a pedestrian by a bicyclist).

# 3 Methodology

In this section, the formulation of the proposed micro-economics micro-mobility Prospect Theory based modeling framework is presented first. The specificities of the model when it relates to 1) pedestrians walking behavior and 2) bicycles dynamics characteristics are offered. Translating this comprehensive model into a simulation process is shown afterward.

## 3.1 Prospect theory model

Following the original Prospect Theory (PT) model *(38)*, we consider that decision makers on shared roadways or spaces have a utility associated with the direction of movement to adopt - defined by an angle $\theta$ - and the speed of movement to choose from - defined by a velocity term $v$. We temporarily do not define subscripts for these terms for a clearer presentation. The resulting $U(v, \theta)$ has two components: *1)* a component related to not being involved in a contact or collision with another decision-maker or object; this component is represented by $U_{PT}$ (the subjective value function). *2)* The second component is associated with being involved in a contact or collision and is mainly governed by three variables: $p_i$, $w_c$ and $k(.)$ that denote the collision probability, collision weighting parameter, and collision seriousness term (that may be a function of other variables), respectively. The decision maker(s) then has(have) to choose his/her/their velocity and direction in order to maximize his/her/their total utility:

$$U(v, \theta) = \{(1 - p_i)U_{PT}(v, \theta) - p_i w_c k(.)\} \qquad (1)$$

The key and most complicated term to define in the above equation is the expected collision probability $p_i$ between two decision makers or a decision maker *i* and an object *j* in a 2-D space. This term was formulated in a simplistic manner in (*39*) for pedestrian-to-pedestrian interaction modeling; however, such formulation does not account for different speeds between objects and thus does not put enough focus on the collision intensity.

In this paper, the collision probability is calculated based on the probability of the Euclidian distance between the two entities being less than a threshold. Let's assume that entity *j* is another decision-maker or an object (for example, a pedestrian or bicyclist, a wall …etc.) and that the locations of decision maker *i* and entity *j* at any time *t* to be given by the coordinates ($x_i$, $y_i$) and ($x_j$, $y_j$) respectively. The collision probability $P_{ij}$ can be written as:

$$P_{ij}^{\square}(\sqrt{(x_i(t) - x_j(t))^2 + (y_i(t) - y_j(t))^2} \leq \Theta) \qquad (2)$$

In this scenario, we are considering the center coordinate of a given entity and the thresholds depends on the type of collision being considered.

At the next time step $t_1 = (t + \Delta t)$, the position of the decision maker $i$ will be defined by the coordinates $(x_i(t + \Delta t), y_i(t + \Delta t))$:

$$x_i(t_1) = x_i(t + \Delta t) = x_i(t) + (v_i \cos \theta_i). \Delta t \qquad (3)$$
$$y_i(t_1) = y_i(t + \Delta t) = y_i(t) + (v_i \sin \theta_i). \Delta t \qquad (4)$$

Where, $v_i$ is the velocity of the decision maker $i$ at time $t$ and $\theta_i$ is the angle associated with the decision maker at time $t$. If the velocity ($v_j$) and the direction ($\theta_j$) of any other decision maker $j$ or an object are assumed to be normally distributed by the decision maker $i$, then:

$$\hat{x}_j(t + \Delta t) = x_j(t) + \hat{v}_{jH}. \Delta t = x_j(t) + N(v_{jH}^c, \sqrt{(\alpha_v^2 v_{jH}^c{}^2)}) \Delta t = x_j(t) + N\left(v_{jH}^c \Delta t, \Delta t^2 \sqrt{(\alpha_v^2 v_{jH}^c{}^2)}\right) =$$
$$N\left((x_j(t) + v_{jH}^c \Delta t), \Delta t^2 \sqrt{(\alpha_v^2 v_{jH}^c{}^2)}\right) = N_1 \qquad (5)$$

And

$$\hat{y}_j(t + \Delta t) = y(t) + \hat{v}_{jV}. \Delta t = y(t) + N(v_{jV}^c, \sqrt{(\alpha_v^2 v_{jV}^c{}^2)}) \Delta t = y_j(t) + N(v_{jV}^c \Delta t, \Delta t^2 \sqrt{(\alpha_v^2 v_{jV}^c{}^2)}) =$$
$$N((y_j(t) + v_{jV}^c \Delta t), \Delta t^2 \sqrt{(\alpha_v^2 v_{jV}^c{}^2)}) = N_2 \qquad (6)$$

$N(.)$ represents a normal distribution; $\alpha_v^2$ is a constant representing uncertainty with respect to the future position in terms of speed choice; and $v_{jH}^c$ and $v_{jV}^c$ are the horizontal and vertical components of the current velocity, respectively. The normal assumption stems from the hypothesis that a decision maker observing another user or object following a direction and a velocity is most likely to assume that such a user or object will keep the corresponding direction and velocity in the next time-step (i.e., the mean values). For a fixed object $j$, the velocity and the corresponding $\alpha_v^{\square}$ are set to zero.

Given the above framework, at time step $t_1$, the collision probability will be:

$$P_{ij}^{\square}(\sqrt{(x_i(t_1) - N_1)^2 + (y_i(t_1) - N_2)^2} \leq \Theta) \qquad (7)$$

Or

$$P_{ij}^{\square}(\sqrt{(N_3)^2 + (N_4)^2} \leq \Theta) \qquad (8)$$

Where, $N_3$ and $N_4$ are two standard normal distribution with

$$Z_1 = \frac{X_j(t) - (x_i(t) - x_j(t) - v_{jH}^c \Delta t)}{\sqrt{(\Delta t^2 \sqrt{(\alpha_v^2 v_{jH}^c{}^2)})}} \qquad (9)$$

$$Z_2 = \frac{Y_j(t) - (y_i(t) - y_j(t) - v_{jV}^c \Delta t)}{\sqrt{(\Delta t^2 \sqrt{(\alpha_v^2 v_{jV}^c{}^2)})}} \qquad (10)$$

For other decision makers, the normal distributions of equation 7 become a chi distribution with 2 degrees of freedom. The probability density function of $\sqrt{(N_3)^2 + (N_4)^2}$ (Equation 8) becomes a chi distribution with 2 degree of freedom and can be written as:

$$\chi(x, 2) = \begin{cases} \dfrac{x^{\square} e^{-\frac{x^2}{2}}}{\Gamma(1)}, & x \geq 0; \\ 0, & otherwise \end{cases} \qquad (11)$$

Where $\Gamma(.)$ represents the gamma function and $x = \sqrt{(N_3)^2 + (N_4)^2}$, for which we seek the probability density. In other words, Equation 8 becomes:

$$P_{ij}^{\square}(\chi(x) \leq \Theta) \qquad (12)$$

The general shape of a chi distribution with 2 degrees of freedom is shown in figure 1. The chi distribution has the same shape: depending on the value of non-centrality parameter, the shape will shift left or right. From figure 1, the shape of the chi distribution with 2 degree of freedom is similar to a slightly right skewed normal distribution. For the simplicity of the implementation purposes and given the possible truncation of the right tail of the distribution, the collision probability will be approximated as a normal distribution.

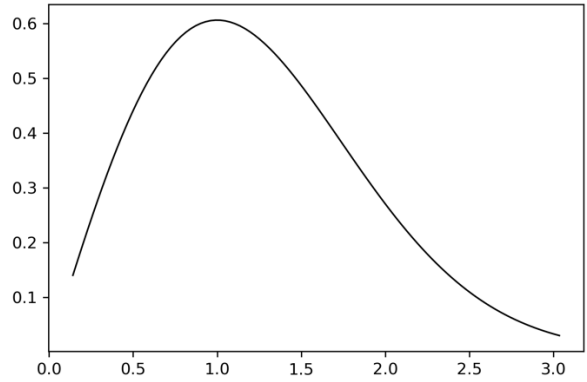

Figure 1: Chi-Distribution with 2 degrees of freedom.

Having different collision possibilities with different surrounding objects or entities, there are two approaches to compute a resulting single collision probability: a) a joint probability distribution may be found; b) a myopic approach may be adopted where the maximum collision probability is computed to find the resulting utility function that governs the future movement of the decision maker. In this study, instead of using the joint probability for the total collision probability (i.e., approach (a)), we will use the maximum of the individual probabilities for each point in the space as the collision probability at that point. In other words,

$$p_i^{\square} = \max\left(P_{i,j}^{\square}\right) \quad \forall j \qquad (13)$$

The second term to define in Equation 1 is the $U_{PT}^{\square}(v, \theta)$ that is the PT value function. This term is mostly associated with the gain experienced with the increase in velocity and the proximity to the desired (predetermined) destination. The angle with respect to the desired destination is defined by $\theta$ and the velocity is defined with respect to a desired velocity. Accordingly:

$$U_{PT}(v_i, \theta_i) = S_\theta [\eta_i \cos(\theta_i)] S_v \left[ \frac{\left(\frac{v_i}{v_{d,i}}\right)}{\left(1+\left(\frac{v_i}{v_{d,i}}\right)\right)^{\frac{\xi_i-1}{2}}} \right] \quad (14)$$

Where $i$ indicates a given decision maker, $\theta_i$ is the angle between the line connecting the decision maker to the destination and the direction of travel, and $v_{d,i}$ is the desired speed for decision maker $i$. $\eta_i$ and $\xi_i$ are parameters to be calibrated: $\eta_i$ is an amplitude parameter that represents the sensitivity to the choice of direction and $\xi_i$ is a measure of non-linearity in the Prospect Theory based value function (PT value function). Two scaling parameters for the angular and velocity dimensions ($S_\theta$ and $S_v$ respectively) are added to translate the utility field into an anticipation field for future decision making.

In prospect theory, the reference point is a key element of the decision-making process. It is the point from which gains and losses are evaluated, and it can influence the perception of the decision outcome. In this model, the reference point for each decision maker at each time step is the current position. Gain or loss is defined by the gain in position towards the desired travel direction. Figure 2 shows the PT value functions value for different velocity and angle. In this figure, the point (0,0) refers to the current position or the reference point of a decision maker. In this illustration, without considering contact/collision with surrounding objects, a decision maker will most likely choose the angle $\theta_i = 0\ rad$ and the velocity $v_i = 1.2\ m/s$. These two terms lead to the highest subjective utility value of 0.4. Note that the choice is not definitive (and thus the wording "most likely") as this model is stochastic in nature as deterministic models may not be the best representation of decision-making processes (which are inherently stochastic). It should be also noted that the value function will change significantly if the reference point at a current time-step is different than ($\theta_i = 0$. $v_i = 0$).

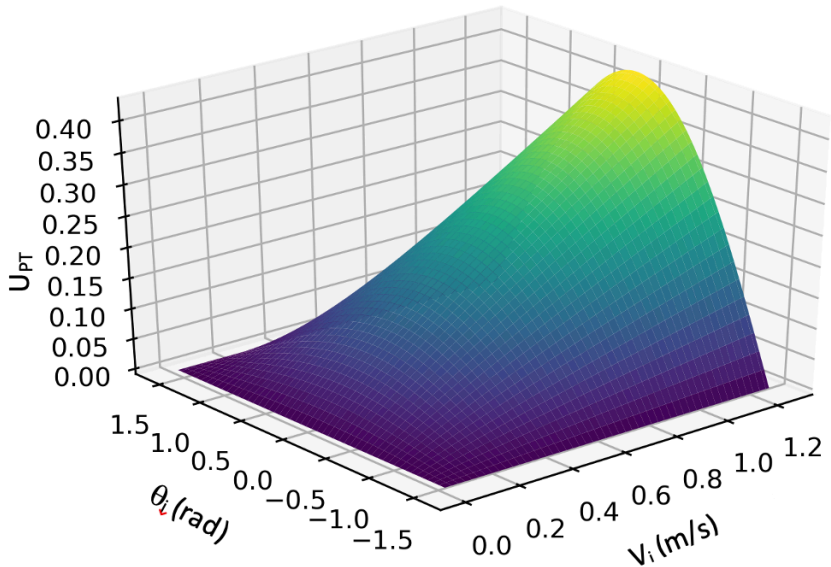

Figure 2: PT value function with respect to reference point ($\theta_i = 0.\ v_i = 0$) at a given time step.

### 3.1.1 Mean and variance of the collision probability function:
Once the distribution characteristics are studied, the collision probability function of Equation 12 is considered as a bivariate normal distribution as follows:

$$P_{i,j} = \frac{1}{2\pi\sigma_1\sigma_2\sqrt{1-\rho^2}} e^{\frac{-Z}{2(1-\rho^2)}} \qquad (15)$$

Where:

$$\rho = \frac{COV(v_{jH}, v_{jV})}{\sigma_1 \sigma_2} \qquad (16)$$
$$\sigma_1 = \sqrt{(\Delta t^2 \sqrt{(\alpha_H^2 v_H^2)})} \qquad (17)$$
$$\sigma_2 = \sqrt{(\Delta t^2 \sqrt{(\alpha_V^2 v_V^2)})} \qquad (18)$$

And $Z = \frac{(v_j^H - \mu_1^2)}{\sigma_1^2} - \frac{2\rho(v_j^V - \mu_1^{\square})(v_j^H - \mu_2^{\square})}{\sigma_1 \sigma_2} + \frac{(v_j^H - \mu_1^2)}{\sigma_2^2}$ (19)

With the mean:
$$\boldsymbol{\mu} = \begin{bmatrix} \mu_1 \\ \mu_2 \end{bmatrix} = \begin{bmatrix} x_j(t) + v_{jH}\Delta t \\ y_j(t) + v_{jV}\Delta t \end{bmatrix} \qquad (20)$$

And the variance:
$$\boldsymbol{\sigma^2} = \begin{bmatrix} \sigma_1^2 \\ \sigma_2^2 \end{bmatrix} = \begin{bmatrix} \Delta t^2 \sqrt{(\alpha_H^2 v_H^2)} \\ \Delta t^2 \sqrt{(\alpha_V^2 v_V^2)} \end{bmatrix} \qquad (21)$$

Now, setting $k(.)$, the collision seriousness term to one we can rewrite Equation 1 as:

$$U = U_{PT}^{\square} - p_i(U_{PT}^{\square} + w_c) \qquad (22)$$

Replacing the collision probability as shown in Equation 15, the mean and the variance covariance matrix of the resulting joint distribution can be calculated as:

$$\mu_U = \left( \begin{bmatrix} \left[ S_\theta[\eta_i \cos(\theta_i)] S_v \left[ \frac{\left(\frac{v_i}{v_{d,i}}\right)}{\left(1+\left(\frac{v_i}{v_{d,i}}\right)\right)^{\frac{\xi_i-1}{2}}} \right] + w_c \right] * \mu_1 + S_\theta[\eta_i \cos(\theta_i)] S_v \left[ \frac{\left(\frac{v_i}{v_{d,i}}\right)}{\left(1+\left(\frac{v_i}{v_{d,i}}\right)\right)^{\frac{\xi_i-1}{2}}} \right] \\ \left[ S_\theta[\eta_i \cos(\theta_i)] S_v \left[ \frac{\left(\frac{v_i}{v_{d,i}}\right)}{\left(1+\left(\frac{v_i}{v_{d,i}}\right)\right)^{\frac{\xi_i-1}{2}}} \right] + w_c \right] * \mu_2 + S_\theta[\eta_i \cos(\theta_i)] S_v \left[ \frac{\left(\frac{v_i}{v_{d,i}}\right)}{\left(1+\left(\frac{v_i}{v_{d,i}}\right)\right)^{\frac{\xi_i-1}{2}}} \right] \end{bmatrix} \right) \qquad (23)$$

$$COV_U = \left[ \left[ S_\theta[\eta_i \cos(\theta_i)] S_v \left[ \frac{\left(\frac{v_i}{v_{d,i}}\right)}{\left(1+\left(\frac{v_i}{v_{d,i}}\right)\right)^{\frac{\xi_i-1}{2}}} \right] + w_c \right] \right]^2 * COV(v_{jH}, v_{jV}) \qquad (24)$$

Equations 23 and 24 are essential for the numerical simulation translating the decision-making processes explained in this section into microscopic trajectories to be shown in the following sections.

## 3.2 Pedestrian model specification and spatial value function analysis

At this stage of the research, for pedestrians, Θ in Equation 12 is considered to be the summation of the 1/2 widths of entities under consideration for collisions between two decision makers (the width may be a shoulder width for other pedestrians depending on the direction of movement or bicycle width/length depending also on the direction of movement). It is equal to only 1/2 width of the decision maker entity for collisions between a pedestrian decision maker and a fixed object. The $U_{PT}(v, \theta)$ that is the PT value function for the (pedestrian) decision makers have been shown in Equation 14 along with a numerical representation of such function in figure 3. For added illustration purposes, we consider the parameters for pedestrian-to-pedestrian interactions: $\eta = 1$, $v_d = 1.5 \, m/s$, $k = 1$, $S_\theta = 1$ and $S_v = 1$. The base scenario consists of a subject pedestrian $i$ traveling directly ahead (left to right with a desired destination to the right) from a (0,0) location. With the aforementioned parametric values, the PT value function, $U_{PT}$ is evaluated starting from simulation time $t = 0$ over a space surrounding the subject pedestrians (figure 3). From figure 3, we can observe that the maximum values of $U_{PT}$ for given $x$ coordinate values are directly ahead of the subject pedestrian location. This is expected since the desired direction of the subject pedestrian is directly ahead of the current position. However, with the increase in non-linearity parameter $\xi$, the changes in $U_{PT}$ values over the space showcases a more myopic perception (i.e., attention to a closer destination location) with less uncertainty (i.e., smaller area characterized by high utility functions shown in darker red).

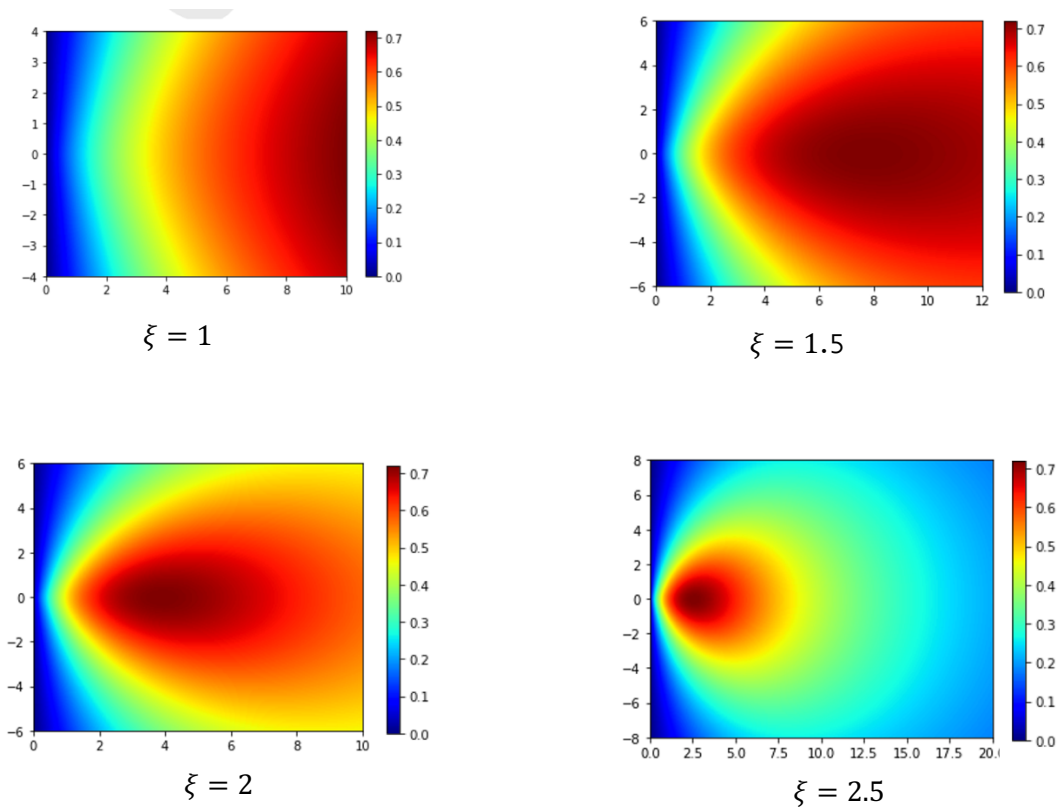

Figure 3: Parametric sensitivity analysis applied on the pedestrian model Value Function $U_{PT}$ (Equation 14) with different ξ values ( $\eta = 1$, $v_d = 1.5 \, m/s$, $k = 1$, $S_\theta = 1$ and $S_v = 1$).

Figure 4a shows the shape of the (subjective) utility function with respect to speed $v_i$ while fixing $\theta_i$ at 0 . This inverted S-shape representation is similar to that seen in Kahnemann and Tversky's work *(40)*. The figure illustrates a more sensitive perception to gains and losses next to the reference point (point of inversion) represented by steeper slopes and a lesser sensitivity to such gains and losses next to the extremes/boundaries of the function. On the other hand, more intuitively, as the speed increases, the associated subjective utility function increases and thus the perception of gaining. Figure 4b shows the generalized utility term ($U(v, \theta)$, Equation 1) with different speed versus spacing terms with consideration of the collision probability. If the spacing is considered along with the collision possibility with another object, there is a balancing exercise between the gains illustrated in figure 4a and the losses associated with colliding with another particle/decision maker. For example, for a speed of 1 m/s, a pedestrian has a low utility associated with small spacing - given the possibility of colliding - and a low utility associated with large spacing - given the loss in not utilizing the space ahead. Accordingly, the highest utilities are seen for spacings between 1 and 1.5 m. Also note that there is a left skewness in the speed distribution with respect to a given spacing value showing a higher tendency to choose lower speeds and thus the risk averse nature of this utility function.

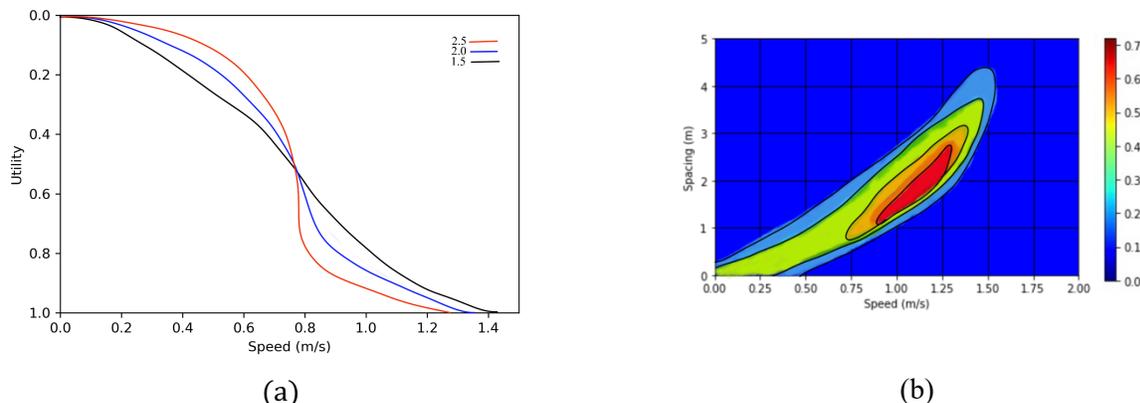

(a)          (b)

Figure 4: (a) Subjective utility vs speed for different values of ξ; (b) Total utility for spacing vs speed terms for pedestrian model.

As a synthesis, with the proper choice of parameters, the offered modeling framework allows explanatory analysis of pedestrians' behaviors. These parameters may be seen in the Table 1. For added illustration, fixing the spacing at 1.5m, the desired speed at 1.5 m/s, and $\xi = 2$, figure 5 shows the subjective utility function (figure 5a), the collision probability (figure 5b) and the total utility function (figure 5c) for three values of collision weights (pedestrian to pedestrian interactions): 160, 200, and 240.

Table 1: Pedestrian (i.e., for pedestrians as decision makers) model parameters and corresponding symbols for the simulation exercise.

| Parameter | Symbols |
|---|---|
| Collision weight with another Pedestrain | $w_{c\ ped}$ (Wc ped: for pedestrian to pedestrian collision) |
| Collision weight with Bicyclist | $w_{c\ ped-bike}$ (Wc ped-bike: for pedestrian to bicycle collision) |
| Collision weight with any fixed object or boundary in the simulation | $w_{c\ ped-barrier}$ (Wc ped-barrier: for pedestrian to barrier/fixed object collision) |
| Amplitude parameter that represents the sensitivity to the choice of direction | η |
| Non-linearity parameter | ξ |
| Reaction time | τ |

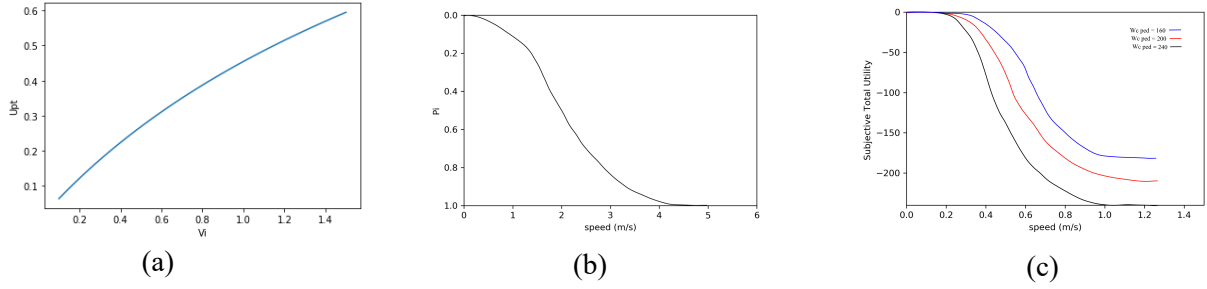

(a)              (b)              (c)

Figure 5: Impact of speed on prospect theory subjective utility, collision probability, and subjective total utility for different $w_c$ values for the pedestrian model.

Since collision weight has no impact on the prospect theory subjective utility and the collision probability, for all three-collision weights, the plots remain the same in figures 5a and 5b. However, the total subjective utility depends on the collision weight values. For the same spacing, lower values of collision weight yield higher speeds while following another pedestrian.

## 3.3 Bicycle model specification and spatial value function analysis

For bicycles, an additional layer of vehicle dynamics needs to be incorporated into the modeling framework to take into account the relationship between the turning angle and the velocity. Towards addressing this issue, Figure 6 illustrates a bicycle or an performing a turn.

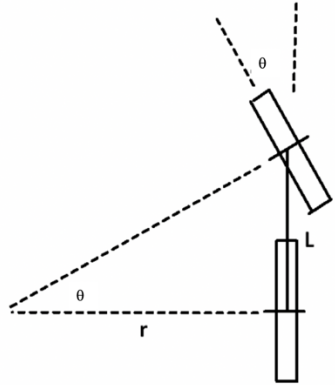

Figure 6: An illustration of a bicycle performing a turn.

Let's assume the turning angle is $\theta$, the wheelbase of the bicycle is $L$ and the turning radius is $r$. Accordingly:

$$\tan\theta = \frac{L}{r} \tag{25}$$

If the lean angle of the cyclist is $\alpha$, then:

$$\tan\alpha = \frac{v^2}{gr} \tag{26}$$

Where $g$ is acceleration due to gravity and $v$ is the speed of the bicycle. Eliminating $r$ from Equation 26, we find:

$$\theta = \tan^{-1}\left(\frac{gL\tan\alpha}{v^2}\right) \tag{27}$$

For the bicycle, the formulation framework in terms of the utility function and the collision probability will be the same as that of the pedestrian model framework. The main difference will be related to the PT value function (Equation 14). The new form of the PT value function becomes:

$$U_{PT}(v_i, \theta_i) = \begin{pmatrix} S_\theta[\eta_i \cos(\theta_i)] S_v \left[ \dfrac{\left(\dfrac{v_i}{v_{d,i}}\right)}{\left(1+\left(\dfrac{v_i}{v_{d,i}}\right)\right)^{\frac{\xi_i-1}{2}}} \right] & for\ \theta_i \leq \left(\tan^{-1}(\dfrac{gL \tan \alpha}{v_i^2})\right) \\ 0 & for\ \theta_i > \left(\tan^{-1}(\dfrac{gL \tan \alpha}{v_i^2})\right) \end{pmatrix} \quad (28)$$

The threshold Θ in the collision probability in Equation 2 will also change accordingly. In this formulation, we will be assuming bicycles as rectangles. Depending on the location of the other object in the simulation scenario, the threshold for collision can be either half of the length or width of the bicycle rectangle. The rest of the formulation of the PT model for the bicycle remains the same as the general Prospect theory-based model presented in section 3.1.

A parametric sensitivity analysis showcasing the properties and the performance of the proposed bicycle model is shown in figure 7. For illustration purposes, we consider the same parameters utilized for pedestrian-to-pedestrian interactions, i.e., $\eta = 1$, $v_d = 5.5\ m/s$, $k = 1$, $S_\theta = 1$ and $S_v = 1$. The base scenario consists of a subject bicyclist $i$ traveling directly ahead (left to right with a desired destination to the right) from (0,0) location. With the aforementioned parametric values, the PT value function, $U_{PT}$ is evaluated starting from simulation time $t = 0$ over an area surrounding the subject bicyclist (figure 7). From figure 7, we can observe that the maximum values of $U_{PT}$ for given $x$ coordinate values are directly ahead of the subject cyclist location as was the case with the pedestrian model. This is expected since the desired direction of the subject decision-maker remains directly ahead of the current position. The change in non-linearity parameter $\xi$ causes a change in $U_{PT}$ values over the space that are also shown in the figure 7. As the non-linearity parameter increases in value, more myopic perception with less uncertainty is still observed (as in figure 3). However, the corresponding patterns are different between figures 3 and 7 given the added layer of dynamics for bicycles.

Figure 8a shows the shape of the utility function with respect to different $\xi$. Figure 8b shows the total utility for different speed versus spacing combinations with consideration of the collision probability. Similar to the pedestrian model, if the spacing is considered along with the collision possibility with another object, there is a balancing exercise between the gains illustrated in figure 8a and the losses associated with colliding with another object/decision maker. For the bicycle model, the highest utilities are seen for spacings between 2.7 and 4.2 m. Also note that there is left skewness in the speed distribution with respect to a given spacing value showing a higher tendency to choose lower speeds and thus the risk averse nature of this utility function.

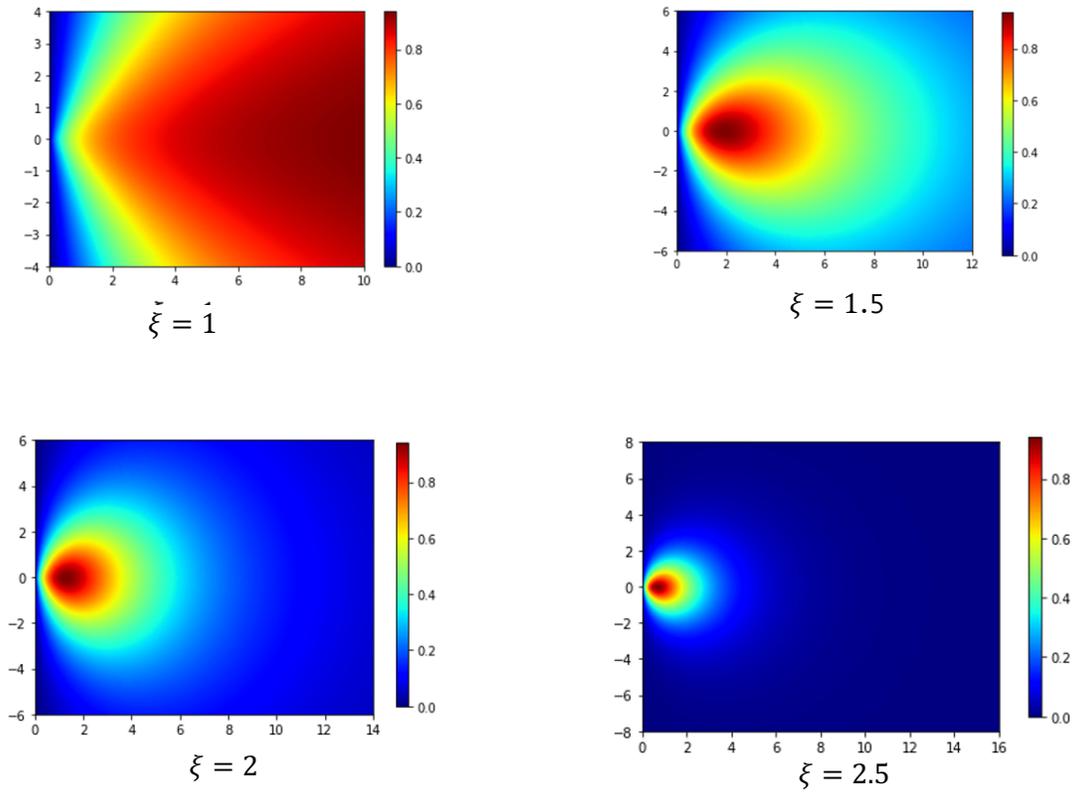

Figure 7: Parametric sensitivity analysis applied on the bicycle model value function $U_{PT}$ (Equation 28) with different $\xi$ values.

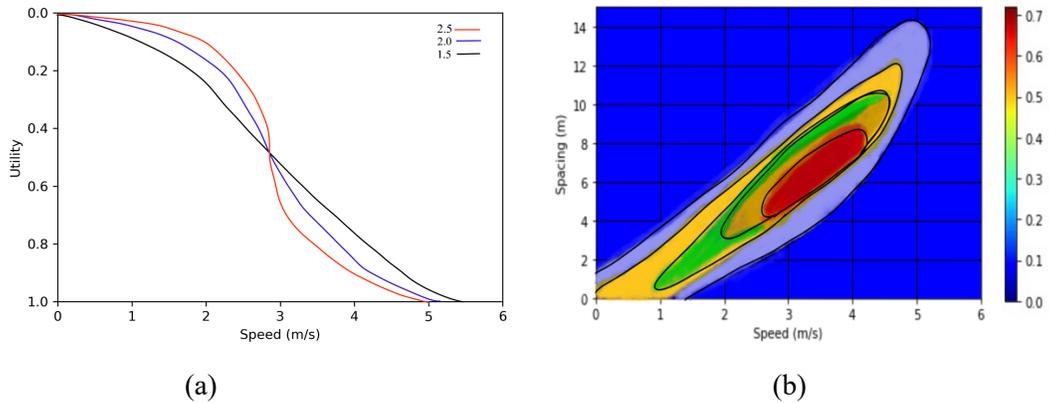

(a)                      (b)

Figure 8: (a) Subjective utility vs Speed for different values of $\xi$; (b) Total utility for spacing vs speed terms for Bicycle Model.

As a synthesis, with the proper choice of parameters, the offered modeling framework allows explanatory analysis of bicyclists' behaviors. Table 2 that offers the main parameters of the bicycle model.

Table 2: Bicycle model parameters (i.e., for cyclists as decision makers) and corresponding symbols for the simulation exercise.

| Parameter | Symbols |
| --- | --- |
| Collision weight with another bicyclist | $w_{c\ bike}$ (Wc bike: for bicycle to bicycle collision) |
| Collision weight with pedestrian | $w_{c\ bike} - ped$ (Wc bike-ped: for bicycle to pedestrian collision) |
| Collision weight with any fixed object or boundary in the simulation | $w_{c\ bike-barrier}$ (Wc bike-barrier: for bicyle to barrier/fixed object collision) |
| Amplitude parameter that represents the sensitivity to the choice of direction | $\eta$ |
| Non-linearity parameter | $\xi$ |
| Reaction time | $\tau$ |

Fixing the spacing at 5m, the desired speed at 5.5 m/s, and $\xi = 2$, figure 9 shows the subjective utility function (figure 9a), the collision probability (figure 9b) and the total utility function (figure 9c) for three values of collision weights: 1600, 2000, and 2400. Notice that the weights tested here are inherently different than those seen in figure 5. Similar to the pedestrian model, the collision weight has no impact on the prospect theory subjective utility and the collision probability. Accordingly, for all three-collision weight values, these curves are the same. However, the total subjective utility depends on the collision weight values. For the same spacing, lower values of collision weights yield to higher speeds while following another bicyclist.

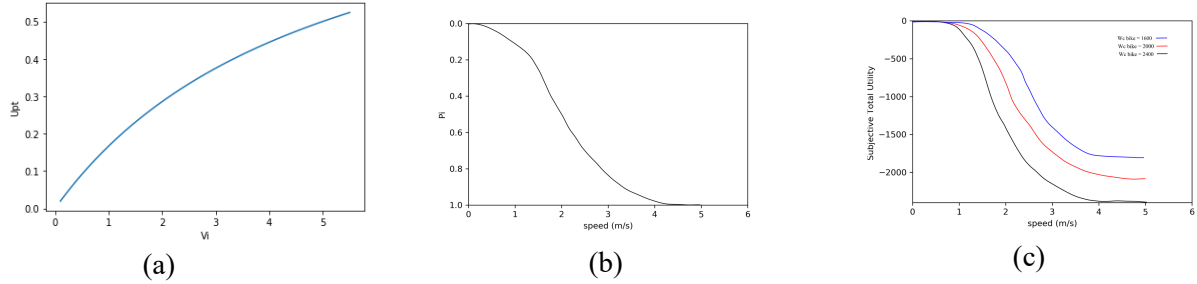

(a)          (b)          (c)

Figure 9: Impact of speed on prospect theory subjective utility (a), collision probability (b), and subjective total utility (c) for different $w_c$ values for the bicyle model.

## 3.4 Approach to numerical simulation

In order to translate the proposed behavioral model into an effective numerical simulation, a Standard Wiener process is applied on both components of the velocity. Towards this goal, the realization of the velocity can be computed as below:

$$v(t) = v^*(t) + \sigma_v(t)\, y_1(t) \qquad (29)$$

Where, $v^*(t)$ denotes the expected value of the velocity and $y_1(t)$ is a Standard Wiener process.

The Standard Wiener process, $y(t)$, is described in a compact way by the following stochastic differential equation:

$$\frac{dy}{dt} = \frac{-y}{\tau} + \xi(t) \qquad (30)$$

Where, $\tau$ is the correlation time and $\xi(t)$ is an uncorrelated stochastic variable (a white noise) defined by its first and second moments:

$$\langle \xi(t) \rangle = 0 \ and \ \langle \xi(t) \xi(t') \rangle = \delta(t - t') \qquad (31)$$

In the above equation, $v_H$ and $v_V$ are defined by the bivariate normal distribution. The mean is shown in Equation 23 and the variance is extracted from Equation 21. The Standard Wiener processes for the velocity components can be written as:

$$v_H(t) = v_H^*(t) + \sigma_{vH}(t) y_1(t) \qquad (32)$$
$$v_V(t) = v_V^*(t) + \sigma_{vV}(t) y_2(t) \qquad (33)$$

The two-dimensional Wiener process can be written as:

$$y(t) = \begin{pmatrix} y_1(t) \\ y_2(t) \end{pmatrix} \qquad (34)$$

Overall, the Standard Wiener process described in Equation 29 has two independent components: $y_1(t)$- the horizontal velocity component - and $y_2(t)$- the vertical velocity component. In this manuscript, the two components of *y(t)* are utilized for numerical computation and analysis.

# 4. Calibration and Validation

This section presents the data used to calibrate the model introduced in the previous section. The calibration is followed by a simulation exercise in order to analyze the microscopic (i.e., trajectories) and macroscopic (i.e., fundamental diagram) properties of the micro-economics modeling framework. Specific attention is given to the behavior of pedestrians and bicyclists at bottlenecks.

## 4.1 Data collection

To calibrate the model described in section 3, we collected data from one location, and we retrieved data from another location - both through video analysis and trajectory extraction. Both trajectory datasets are recorded for a 15-minute duration. From the trajectories extracted, we focus on the pedestrian and bicycle modeling frameworks and their analysis in this paper.

The data is extracted from videos recorded in front of the Lincoln Memorial in Washington DC, USA (Dataset 1) and at a transit station in Amsterdam, the Netherlands (Dataset 2) *(6)*. The videos used for this paper has a 15 min duration with several conflict points detected between pedestrians and cyclists sharing the same space. Snapshots of the videos are shown in figure 10 along with the resulting extracted trajectories. The units of all the trajectory plots presented in this study are in meters.

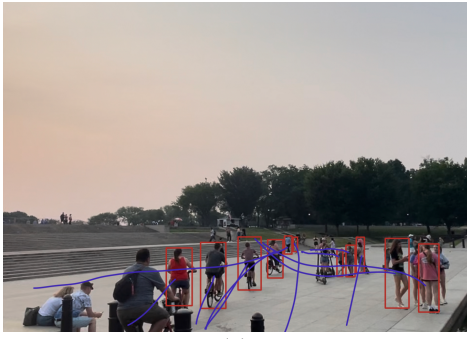
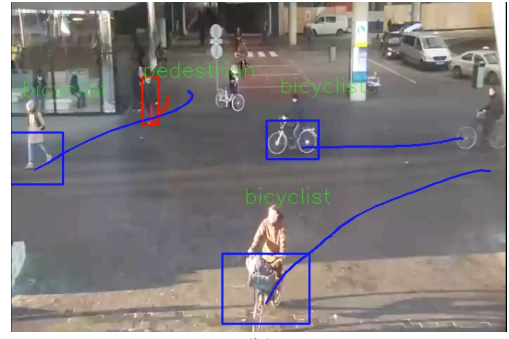
(a) (b)

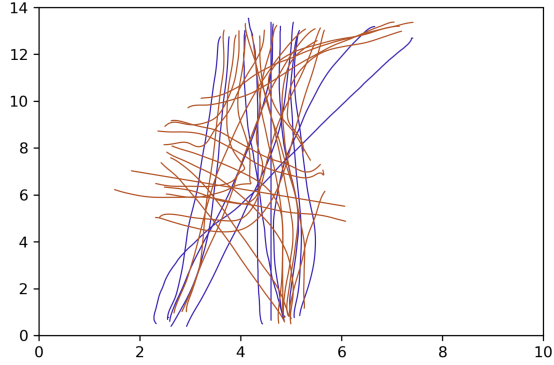
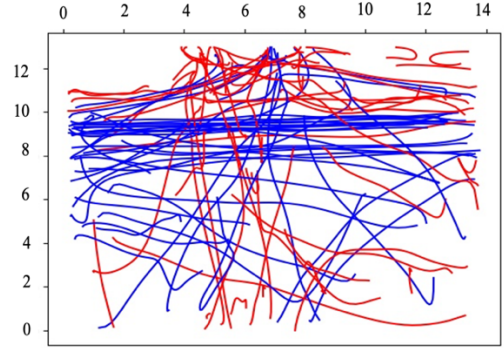
(c) (d)

Dataset 1: Washington DC, USA           Dataset 2: Amsterdam, The Netherlands
Figure 10: Video detection snapshots of Dataset 1 (a) and Dataset 2 (b) and trajectory extraction of pedestrians and bicyclists in Dataset 1 (c) and Dataset 2 (d).

In order to estimate and draw the fundamental diagram based on the trajectory data, space mean speeds and densities were estimated. In the videos, the velocities of the pedestrians and bicyclists have different directions. To calculate the scalar value of the velocity, the following equation is used:

$$v_{space} = \sqrt{v_x^2 + v_y^2} \qquad (35)$$

Where $v_x$ is the scalar projection of the velocity vector in the $x$ direction, and $v_y$ is the scalar projection of the velocity in the $y$ direction. The scalar projections of the velocities (i.e., speeds) were then aggregated into space mean speed measures depending on the study area under consideration and the time duration of interest.

For density calculations, a voronoi-based method *(45)* of measuring density is used in figure 11. Each pedestrian is assigned to a cell within a given area. This area is defined as the sum of the Voronoi areas of the cells located in the analysis area. Each point represents a particular object. (e.g., pedestrian, bicyclist). The method can be defined as:

$$D_v = \frac{N}{\sum_{i=1}^{N} A_i} \qquad (36)$$

Each subject *i*, occupies a cell of area $A_i$. The total area $A$ is the sum of the Voronoi's areas $A_i$ of cells overlapping with the study area. This method is widely used in pedestrian density measurement in literature *(46-47)*. The speed versus (vs.) density relationship of the collected data is shown in figure 11.

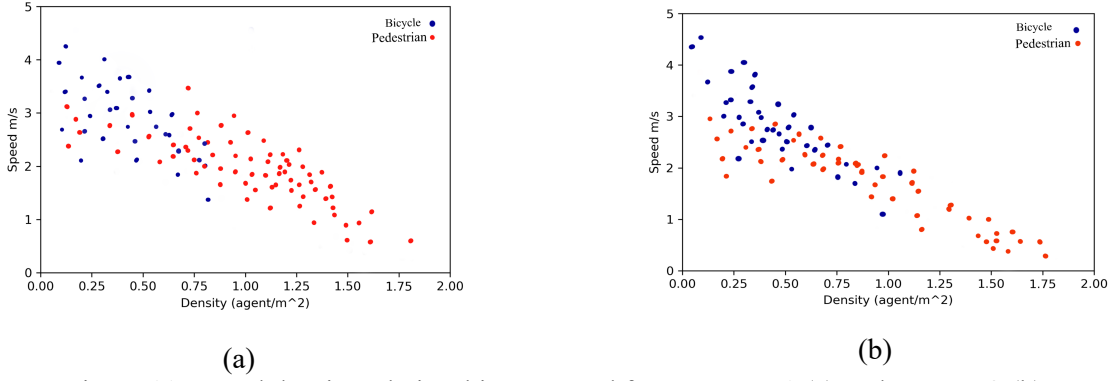

(a)                                        (b)

Figure 11: Speed density relationship extracted from Dataset 1 (a) and Dataset 2 (b).

It should be noted that Dataset 1 is collected from a touristic park area. It thus shows more scattering in the speed density data points in figure 1 (left) - possibly because of arrival of large groups of tourists at some times with irregular behavior associated with sightseeing and taking pictures. Dataset 2 is collected in front of a transit station with the majority of the users being commuters. It shows a speed-density relation in line with the literature *(41-42)*.

## 4.2 Model calibration and validation

This section aims at illustrating the results retrieved when calibrating the microscopic generalized micro-mobility micro-economics based model presented in section 3. Since the model is non-linear and stochastic in nature, the corresponding optimization problem is also nonlinear. For this nonlinear calibration problem, a genetic algorithm (GA) is used to minimize a root mean square error (RMSE) term between the spacing of the recorded trajectories and the simulated trajectories. The RMSE term may be seen in the equation below:

$$RMSE(s) = \sqrt{\frac{1}{T} \sum_{t=1}^{T}[s^{sim}(t) - s^{obs}(t)]^2} \qquad (37)$$

Where, $s^{sim}(t)$ and $s^{obs}(t)$ represent the simulated and actual spacing of the subjects. The GA algorithm will directly compare the trajectories observed to the trajectories obtained from the microscopic simulation while prespecifying the initial conditions and the final destination of travel. The spacing error is calculated at every time step (0.1. second) for a given subject. At the next time step, the future position of the subject is then calculated and compared to recorded position in the present time step while considering the surrounding subjects/obstacles. The final objective function to minimize is the RMSE error (Equation 37) based on the entire observation period. This selection is made because such an error structure is not sensitive to the type of congestion regime a pedestrian/bicyclist is encountering. If we use the velocity difference as the base of the objective function to be minimized, the absolute error may be more sensitive to changes in the empirical data when speeds are higher, and the relative error may be more sensitive to changes at lower speeds *(38-39)*. During each chromosome generation, the RMSE error fitness is calculated, and greedy selection is used to pick the parameters with the 10 best fitness scores. These become "parents" that are then used to generate chromosomes which are combined to create "children". During this combination, a crossover point is chosen using random selection, and genes, except for the chromosome with the single best fitness score, are randomly mutated with a probability rate of 10%. At the beginning, a fixed number of generations are assessed, and the process ends when the fitness score drops below 20 cm or there is no improvement during 15 successive chromosome generations. For the pedestrian modeling framework, the initial set of parametric values used for each parameter are: $W_c = 10, \eta = 1, \xi = 1$ and $\tau = 0.1$. For the bicycle modeling framework, the initial set of parametric values used for each parameter are: $W_c = 100, \eta = 1, \xi = 1$ and $\tau = 0.1$.

These values are chosen based on the sensitivity analysis and verification exercise performed during the formulation presented in section 3. This GA based calibration is applied on both trajectory datasets. The first 10 minutes of the 15 minutes data is used for the calibration of the model.

In order to measure the goodness of fitting, the Relative Root Mean Squared Error (RRMSE) is also calculated using the equation (38).

$$RRMSE(s) = \sqrt{\frac{\frac{1}{T}\sum_{t=1}^{T}[s^{sim}(t)-s^{obs}(t)]^2}{\sum_{t=1}^{T}(s^{obs}(t))^2}} * 100 \qquad (38)$$

According to *(48)* model accuracy is considered excellent when RRMSE is less than 10%, good if 10% to 20%, fair if 20% to 30%, and poor if RRMSE is greater than 30%.

Tables 3, 4 and 5 present the calibration results from the data collected in Washington DC. The RRMSE value of the dataset 1 calibration is 9%, which represents an excellent fit.

Table 3: Calibrated parametric values of the Pedestrian Model – Dataset 1.

| Parameter | Average | Minimum | Maximum |
|---|---|---|---|
| Wc ped | 190 | 175 | 205 |
| Wc ped-bike | 215 | 184 | 315 |
| Wc ped-barrier | 87 | 29 | 114 |
| $\eta$ | 3 | 2.4 | 4.2 |
| $\xi$ | 3.6 | 4.2 | 2.3 |
| $\tau$ (Reaction time) | 1.2 | 0.7 | 1.7 |

Table 4: Calibrated parametric values of the Bicycle Model – Dataset 1.

| Parameter | Average | Minimum | Maximum |
|---|---|---|---|
| Wc bike | 1900 | 1550 | 2200 |
| Wc bike-ped | 1790 | 1560 | 2170 |
| Wc bike-barrier | 1350 | 950 | 1600 |
| $\eta$ | 4.5 | 3.3 | 5.9 |
| $\xi$ | 4.1 | 5.4 | 3.3 |
| $\tau$ (Reaction time) | 0.4 | 0.3 | 0.8 |

Table 5: Average RMSE error statistics – Dataset 1.

| Statistics | Values (cm) |
|---|---|
| Average | 3.3 |
| Minimum | 0.25 |
| Maximum | 16.3 |
| Standard Deviation | 1.89 |

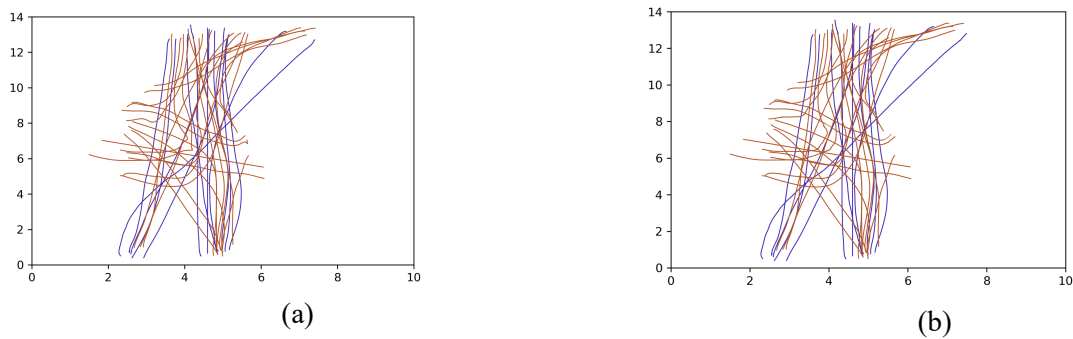

(a)            (b)

Figure 12: (a) Real vs (b) simulated trajectories for Dataset 1.

Figure 12 shows the real vs simulated trajectories collected and extracted in Washington, DC (Dataset 1). The red lines represent the pedestrian trajectories, and the blue lines represent the bicycle trajectories. Since this is a touristic area, some decision-makers were stopping in the data collection zone to watch the monuments in the surrounding area. Accordingly, multiple destination points within the study area were retrieved for the same trajectories and the operational behavior was modeled between these destinations based on the modeling framework presented earlier. This is consistent with the objective of the manuscript as we are offering an operational decision-making model and not a destination choice tactical or strategic decision-making model.

The distributions of the different calibrated parameters values for the Washington DC dataset are illustrated in the figure 13. From the figure, it can be observed that most of the parameters have a Gaussian shaped distribution functions with a governing peak value. For some of the parameters, there is a peak at the lower (Wc ped-bike, $\xi$ (ped), Wc bike-ped) and the upper (Wc ped-barrier) of the distributions that shows a specifc class of behavior related to the type of interactions. Tables 3 and 4 show the minimum, maximum and average values of this distributions.

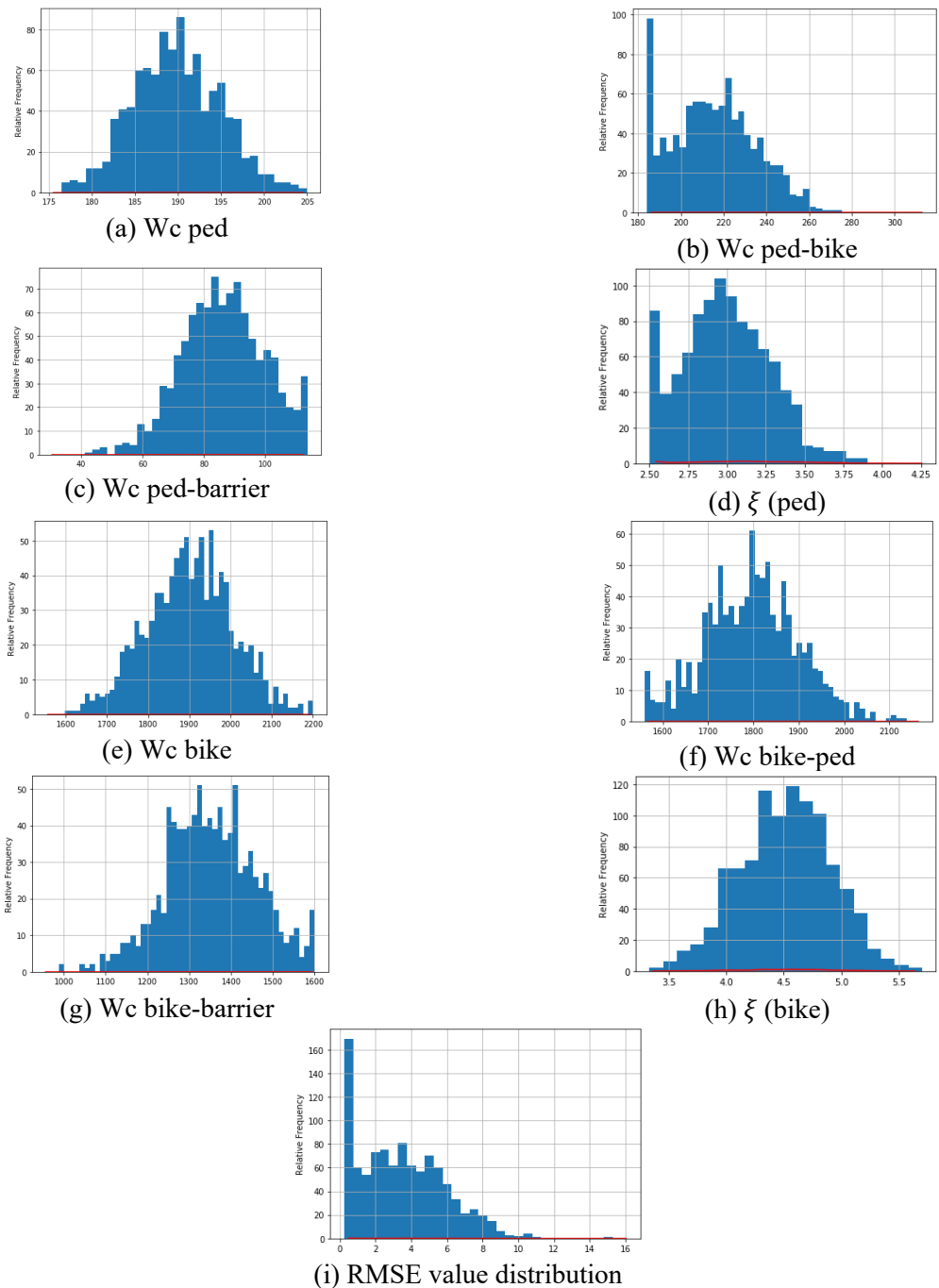

Figure 13: Parametric value (a though h) and error distribution (i) of modeling parameters – Dataset 1.

A correlation analysis is done to check the parametric correlation of the model after the calibration. The correlation matrices shown in tables 6 and 7 show mainly low correlation values between different parameters. This may indicate a high level of independence between parameters which is a desirable property. The highest correlation is associated with the different types of collision weight parameters which is expected. However, such correlation level does not motivate the need of having a single collision weight parameter at this stage as may be seen in the differences between their associated values.

Table 6: Correlation matrix for Pedestrian model's calibrated parameters using GA – Dataset 1.

|  | Wc ped | Wc ped-bike | Wc ped-barrier | $\eta$ | $\xi$ | $\tau$ |
|---|---|---|---|---|---|---|
| Wc ped | 1 | 0.34 | 0.38 | 0.12 | 0.25 | 0.02 |
| Wc ped-bike | 0.34 | 1 | 0.33 | 0.13 | 0.25 | 0.02 |
| Wc ped-barrier | 0.38 | 0.33 | 1 | 0.17 | 0.22 | 0.02 |
| $\eta$ | 0.12 | 0.13 | 0.17 | 1 | 0.13 | 0.04 |
| $\xi$ | 0.25 | 0.25 | 0.22 | 0.13 | 1 | 0.039 |
| $\tau$ | 0.02 | 0.02 | 0.02 | 0.04 | 0.039 | 1 |

Table 7: Correlation matrix for Bicycle model's calibrated parameters using GA – Dataset 1.

|  | Wc bike | Wc bike-ped | Wc bike-barrier | $\eta$ | $\xi$ | $\tau$ |
|---|---|---|---|---|---|---|
| Wc bike | 1 | 0.35 | 0.30 | 0.17 | 0.25 | 0.03 |
| Wc bike-ped | 0.35 | 1 | 0.31 | 0.15 | 0.24 | 0.03 |
| Wc bike-barrier | 0.30 | 0.31 | 1 | 0.12 | 0.21 | 0.04 |
| $\eta$ | 0.17 | 0.15 | 0.12 | 1 | 0.13 | 0.04 |
| $\xi$ | 0.25 | 0.24 | 0.21 | 0.13 | 1 | 0.032 |
| $\tau$ | 0.03 | 0.03 | 0.04 | 0.04 | 0.032 | 1 |

Tables 8, 9 and 10 shows the calibrated results of the data from Amsterdam (Dataset2). The RRMSE value of the dataset 2 calibration is 11%, which represents a good fit. The values of the collision weight parameters are generally higher in this dataset. This first observation may be associated with decision-makers' perception of contact/collision: less concern is exhibited among Washington DC tourists to collide and be in closer proximity to others if compared to commuters in the Netherlands.

Table 8: Calibrated parametric values of the Pedestrian Model – Dataset 2.

| Parameter | Average | Minimum | Maximum |
|---|---|---|---|
| Wc ped | 196 | 179 | 225 |
| Wc ped-bike | 247 | 205 | 346 |
| $\eta$ | 3.5 | 2.6 | 4.7 |
| $\xi$ | 3.6 | 2.3 | 4.2 |
| $\tau$ (Reaction time) | 1.3 | 0.8 | 1.6 |

Table 9: Calibrated parametric values of the Bicycle Model – Dataset 2.

| Parameter | Average | Minimum | Maximum |
|---|---|---|---|
| Wc bike | 2158 | 1785 | 2653 |
| Wc bike-ped | 1995 | 1690 | 2234 |
| $\eta$ | 4.2 | 3.1 | 5.5 |
| $\xi$ | 4.6 | 3.4 | 5.8 |
| $\tau$ (Reaction time) | 0.5 | 0.3 | 0.7 |

Table 10: Average RMSE error statistics– Dataset 2.

| Statistics | Values (cm) |
|---|---|
| Average | 3.5 |
| Minimum | 0.3 |
| Maximum | 16.7 |
| Standard Deviation | 1.7 |

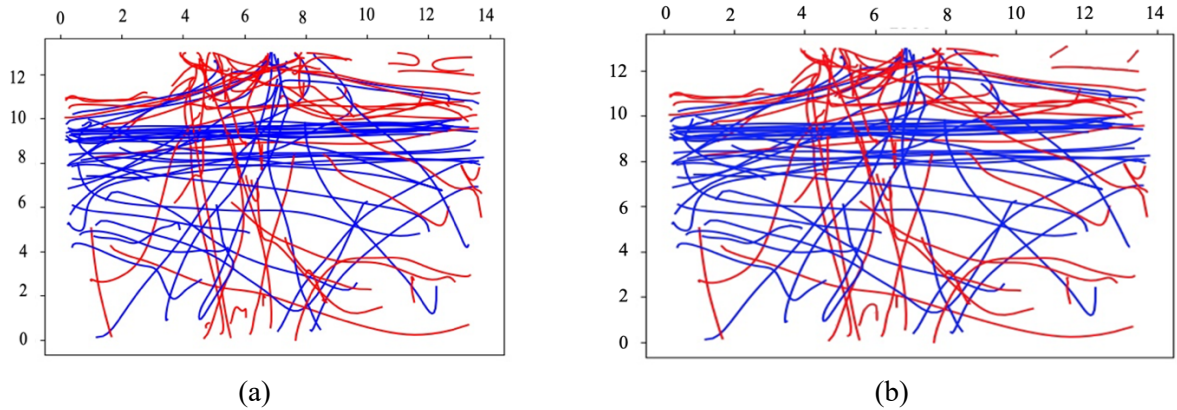

(a)                 (b)

Figure 14: (a) Real vs (b) simulated trajectories for Dataset 2.

Figure 14 shows the simulated versus the real trajectories extracted in Dataset 2. As we mentioned earlier, this dataset is collected at a transit station. The distributions of the different calibrated parameters values for Dataset 2 are illustrated in the figure 15. Table 8 and 9 shows the minimum, maximum and average values of this distributions. The correlation analysis for Dataset 2 is shown in tables 11 and 12. This analysis also shows low level of correlation among the model parameters.

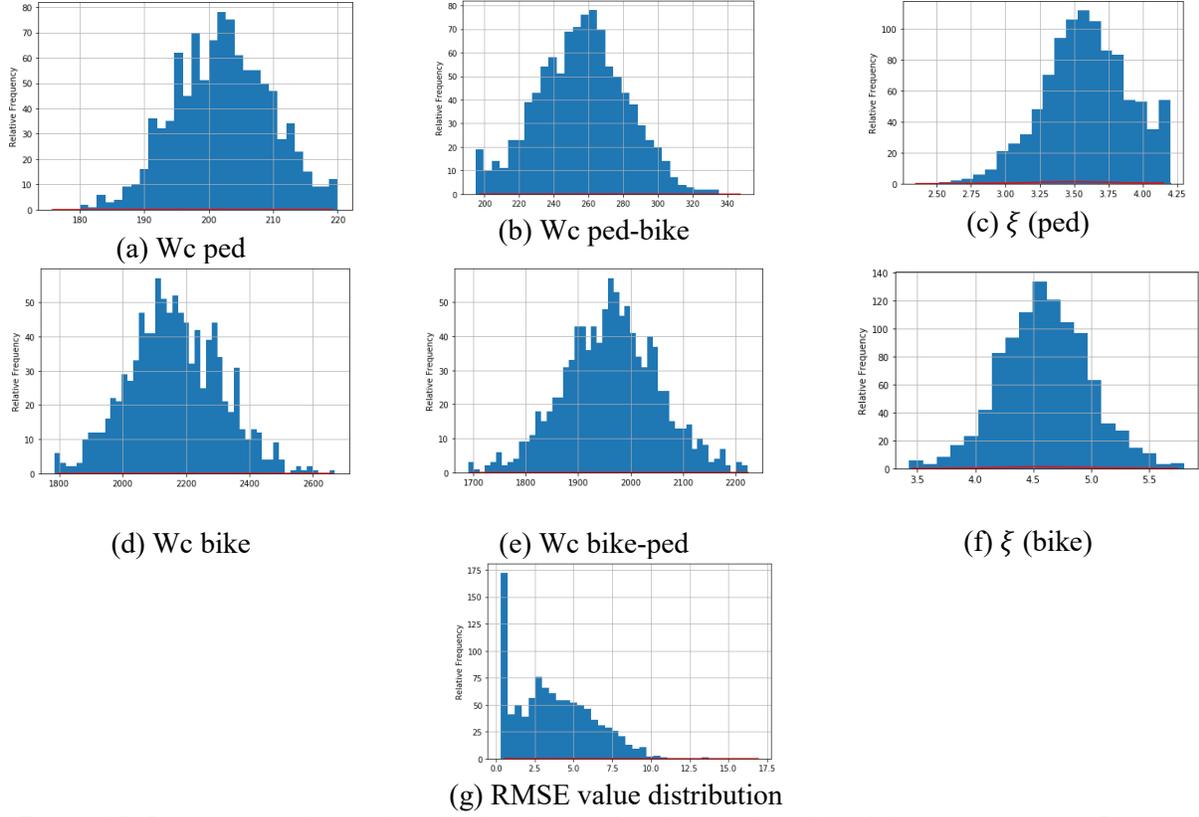

Figure 15: Parametric value (a though f) and error distribution (g) of modeling parameters – Dataset 2.

Table 11: Correlation matrix for Pedestrian model's calibrated parameters using GA – Dataset 2.

|  | Wc ped | Wc ped-bike | $\eta$ | $\xi$ | $\tau$ |
|---|---|---|---|---|---|
| Wc ped | 1 | 0.34 | 0.11 | 0.25 | 0.06 |
| Wc ped-bike | 0.34 | 1 | 0.15 | 0.2 | 0.03 |
| $\eta$ | 0.11 | 0.15 | 1 | 0.16 | 0.04 |
| $\xi$ | 0.25 | 0.2 | 0.16 | 1 | 0.02 |
| $\tau$ | 0.06 | 0.03 | 0.04 | 0.02 | 1 |

Table 12: Correlation matrix for Bicycle model's calibrated parameters using GA – Dataset 2.

|  | Wc bike | Wc bike-ped | $\eta$ | $\xi$ | $\tau$ |
|---|---|---|---|---|---|
| Wc bike | 1 | 0.33 | 0.13 | 0.20 | 0.05 |
| Wc bike-ped | 0.33 | 1 | 0.12 | 0.23 | 0.02 |
| $\eta$ | 0.13 | 0.12 | 1 | 0.13 | 0.04 |
| $\xi$ | 0.20 | 0.15 | 0.13 | 1 | 0.03 |
| $\tau$ | 0.05 | 0.02 | 0.04 | 0.03 | 1 |

The correlation observations are consistent among the two datasets presented in this manuscript. On the other hand, there is an order of magnitude difference between the collision weights associated with pedestrians if compared to the collision weights associated with bicyclists. In other words, pedestrians put less weight on collision if compared to bicyclists in general. Within the pedestrian decision-making process, there is more sensitivity towards colliding with other bicycles if compared to the sensitivity colliding with pedestrians. The least sensitivity is towards colliding with barriers. On the other hand, for the bicyclists' decision-making process, it may be seen that there is less sensitivity towards colliding with pedestrians if compared to that exhibited when colliding with other bicycles. The least sensitivity is towards colliding with barriers. The interplay between the collision weights and the remaining parameters makes this modeling framework suitable for additional risk-based or safety-based analysis of micro-mobility traffic in shared spaces.

To assess the robustness of the parameter values calibrated in this section, the model is validated using the last 5 minutes of Dataset 1 and Dataset 2. The corresponding error statistics are shown below. The validation values are comparable to those seen in the calibration exercise and thus the order of the parameters shows consistency for additional simulation analysis.

Table 13: Average RMSE statistics - Dataset 1.

| Statistics | Values (cm) |
|---|---|
| Average | 3.8 |
| Minimum | 0.4 |
| Maximum | 18.0 |
| Standard Deviation | 2.2 |

Table 14: Average RMSE statistics - Dataset 2.

| Statistics | Values (cm) |
|---|---|
| Average | 3.5 |
| Minimum | 0.3 |
| Maximum | 15.5 |
| Standard Deviation | 1.9 |

Finally, as an added cross-data validation exercise, with the calibrated parametric values of Dataset 1, we simulated Dataset 2. Table 15 below shows the error statistics found.

Table 15: Average RMSE statistics of Dataset 2 simulated with Dataset 1's calibrated parameters.

| Statistics | Values (cm) |
|---|---|
| Average | 3.9 |
| Minimum | 0.35 |
| Maximum | 15.2 |
| Standard Deviation | 1.96 |

The average error went up from 3.8 cm to 3.9 cm. This error range remains reasonable. In conclusion, the presented modeling framework is able to show slight differences in the parameters indicating differences in decision-making processes in different locations. However, the model at hand is still able to reproduce trajectories and microscopic dynamics in a robust manner irrespectively of the parametric values utilized.

# 5 Numerical Analysis

In this section, we compare the model's performance with one of the latest mixed models for pedestrian and bicyclists (*49*). This particular model was tested using the dataset 1 and dataset 2 and benchmarked against the PT based model presented in this study.

Following this benchmarking exercise, we simulate different scenarios with the calibrated parameters of the previous section to further assess the performance of the proposed generalized two-dimensional micro-economics model. The main types of interactions to be assessed are those seen during bottleneck formation, lane formation and shockwave formation. An additional bottleneck scenario for pedestrian interactions is recreated when urgency is assumed as is the case during evacuation conditions.

## 5.1 Cross comparative evaluation

In order to compare the performance of the presented prospect theory (PT) based model with an existing microsimulation model, we have selected the heuristic-based model presented by *(49)*. The acceleration of the pedestrians and bicyclist are described in this model with the following equations.

$$\frac{dv}{dt} = \frac{v_{des} - v}{\tau_1} \quad (39)$$

$$\frac{dv_i}{dt} = \begin{cases} \min\left(\left|\frac{v_{des1} - v_i}{\tau_2}\right|, a_a\right) \cdot \frac{v_{des1}}{|v_{des1}|} + \frac{v_{des2}}{\tau_4}, & |v_{des1}| \geq |v_i| \\ -\min\left(\left|\frac{v_{des1} - v_i}{\tau_3}\right|, a_d\right) \cdot \frac{v_{des1}}{|v_{des1}|} + \frac{v_{des2}}{\tau_4}, & |v_{des1}| < |v_i| \end{cases} \quad (40)$$

Where $v_{des}$ is desired velocity, $v$ is the current velocity $\tau_1$ is relaxation time for pedestrian and $\tau_2$, $\tau_3$ and $\tau_4$ are relaxation times for acceleration, deceleration, and turning for bicyclists respectively, and $a_a$ and $a_d$ are the maximum acceleration and maximum deceleration of the bicycle. The model is calibrated using the same methodology described in section 4.2. The calibrated parametric values of pedestrians and bicyclists for both dataset 1 and dataset 2 are presented in Table 16 and 17.

Table 16: Calibrated parametric values of Pedestrian for the heuristic-based model.

| Parameter | Dataset 1 | Dataset 2 |
|---|---|---|
| $\alpha$ | 13 | 8 |
| $V_{max}$ | 3.5 m/s² | 3 m/s² |
| $\tau_1$ | 0.55 | 0.50 |
| $T_{safe}$ | 0.30 | 0.26 |

Table 17: Calibrated parametric values of Bicyclists for the heuristic-based model.

| Parameter | Dataset 1 | Dataset 2 |
|---|---|---|
| $\alpha$ | 5 | 4 |
| $V_{max}$ | 4.3 m/s² | 4.5 m/s² |
| $\tau_1$ | 0.80 | 0.75 |
| $\tau_2$ | 0.45 | 0.50 |
| $\tau_3$ | .10 | .10 |
| $\tau_4$ | .10 | .10 |
| $T_{safe}$ | 0.25 | 0.20 |

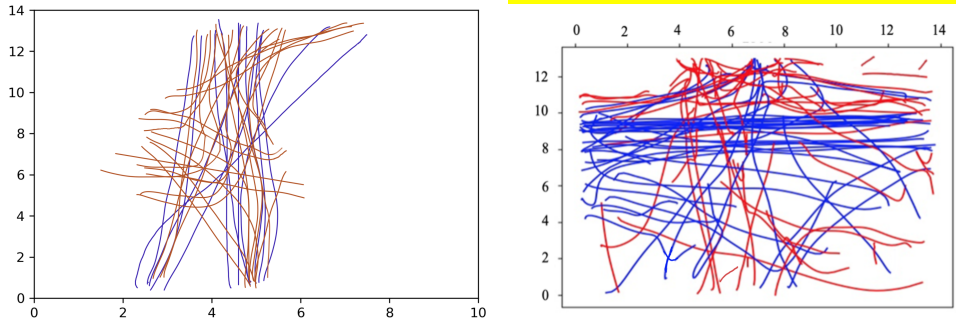

Figure 16: Simulated Trajectories using heuristic based model: (a) dataset 1 (b) dataset 2

Table 18: Average RMSE statistics of the heuristic-based model

| Statistics | Dataset 1 | Dataset 2 |
|---|---|---|
| Average | 4.1 | 3.95 |
| Minimum | 0.45 | 0.41 |
| Maximum | 20.1 | 19.1 |
| Standard Deviation | 2.8 | 2.3 |

Table 18 presents the RMSE statistics of the heuristic-based model. The RMSE values of this heuristic based model is slightly higher than the PT based model (Table 5 and Table 10) for both datasets. The error distribution for both cases have higher minimum, maximum and standard deviation than the PT based model. The RRMSE of dataset 1 and dataset 2 are 11% and 15% respectively. Which is higher than the PT model's RRMSE values for both datasets.

## 5.2 Simulation experiments

### 5.2.1 Bottleneck scenario

We simulated a 10-minute bottleneck formation with the calibrated parameters from Dataset 1. The bottlenecks are generated for pedestrians and bicycles separately for assessment purposes. The figure below shows the trajectories produced. At capacity, the speed for pedestrians is found to be approximately 1 m/s. The maximum density at the entry of the bottleneck is found to be 1.2 ped/m². It should be noted that the maximum density that can be reached with this model is related to the value of the collision weight parameter $w_c$ (Wc). In section 4.4, this relationship is further studied. The results of the simulated pedestrian bottleneck are consistent with the results from literature *(41)*. In case of the bicycle model, the trajectories form a funnel like shape at the bottleneck. The speed at capacity in this analysis is approximately 2.8 m/s. The maximum density at the entry of the bottleneck is found to be 0.76 bicycle/m².

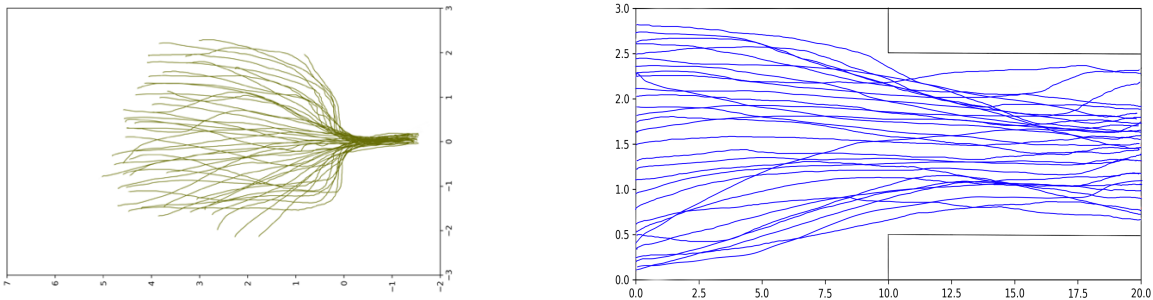

(a) (b)

Figure 17: Pedestrian (a) and bicycle (b) trajectories simulated at a bottleneck.

### 5.2.2 Lane formation scenario – pedestrians

Lane formation is a specific type of self-organized behavior exhibited by pedestrians especially when dealing with two directional flows. Accordingly, we simulated with the calibrated parameter values of Dataset 1 for 10-minutes of interactions. Pedestrians are generated from left and right of the simulation environment at random locations along the left and right edges. The generation time is also random. The generated trajectories are shown in figure 18(a) (left). Figure 18(b) (right) shows a snapshot at the timestep 500 (50 seconds). The model does not consistently form elaborate lanes that are observed in empirical experiments (*42*) for bidirectional pedestrian flow. However, herding behavior is seen that lead to the formation of lanes and clusters of directional movements as clearly seen in figure 18.

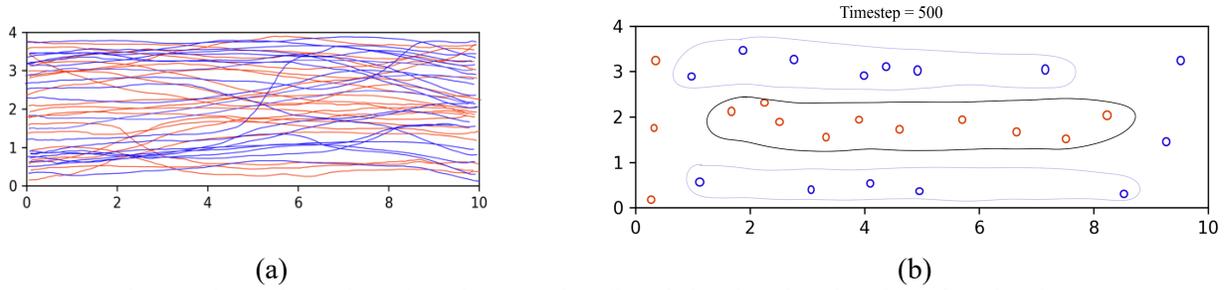

(a)                                      (b)
Figure 18: Trajectories (a) and a snapshot (b) of simulated pedestrians forming lanes.

### 5.2.3 Exit with urgency: pedestrian evacuation

We simulated an evacuation scenario with a narrow bottleneck for pedestrians to explore the possibility of observing some dynamics already reported in the literature. Figure 19 shows the trajectories at the entrance of the bottleneck where we forced a large number of pedestrians generated simultaneously to move towards the same destination/exit. From the trajectories and the density heatmap reported, a cluster of pedestrians approaches the exit point together (figure 19 – bottom) until forming a semi-circular pattern (top) at the bottleneck walls. This behavior is different that the funneling behavior reported earlier and is already reported in the literature (*43*). The semi-circular trajectory shape indicates a high pressure at the bottleneck wall (straight vertical line of the semi-circle which is a precursor of stampede and lateral movement especially at the center of the semi-circle/bottleneck).

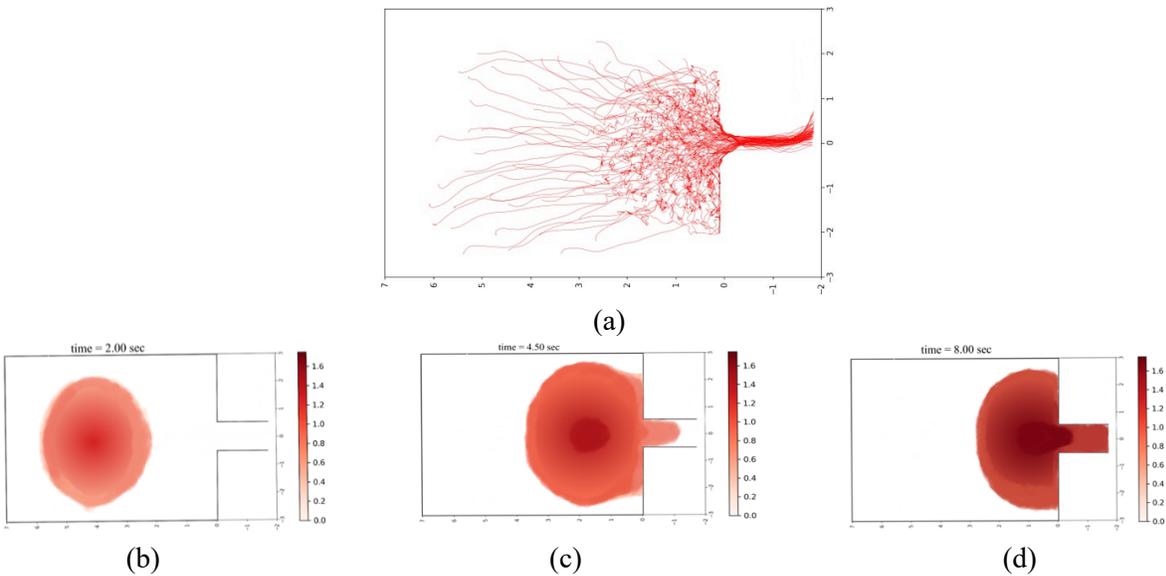

Figure 19: Evacuation scenario trajectories (a) and density heatmaps for pedestrians at simulation times 2.00 second (b), 4.50 second (c) and 8.00 second (d).

### 5.2.4 Emergence and propagation of shockwave

To study the shockwaves of pedestrians and bicyclists, we have simulated them in ring-shaped tracks. For pedestrians, the ring shape track is 1 m wide; the inner radius of the track is 8m. For the bicycles, the track is 3 m wide, and the inner radius is 8m. the whole ring-shaped track is segmented in 8 subareas for the convenience of the shockwave analysis. Each 45-degree arc of the circular track defines a subarea of the track. For the case of pedestrians, 100 pedestrians are simulated using the calibrated parameters of the dataset 1 and the desired velocity set to 1 m/s. At the beginning of the simulations, the pedestrians are generated randomly in space in the track and start moving counterclockwise. For the bicycles, 100 bicyclists were simulated using the dataset 1 calibrated parameter values and the desired velocity set to 4 m/s. The simulated density evolution of pedestrians' and bicyclists is shown in figure 20.

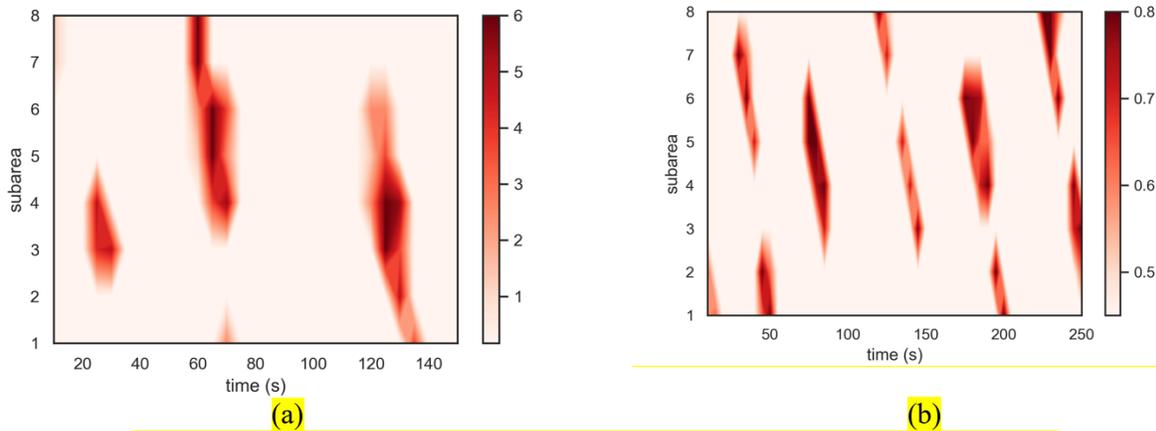

(a)         (b)

Figure 20: Density Evolution at 8 subareas for (a) pedestrians and (b) bicyclists

As shown in figure 20 the stop-and-go waves spontaneously emerged in both simulations. The density in the jam was approximately $\rho_1 = 6$ pedestrian/m$^2$. The density in the moving flow area was approximately $\rho_2 = 1.2$ pedestrian/m$^2$, and the flow rate was $q = 0.96$ pedestrian/s/m. The wave propagation speed can be calculated by $q/(\rho_1 - \rho_2)$ and yields a propagation speed of 0.20 m/s. *(51)* did a shockwave propagation analysis in which the range of wave propagation speed for a desired velocity of 1 m/s is found to be among 0.15 m/s to 0.18 m/s. Our simulated result is similar to the values reported in the referenced paper. For the bicycle simulation, the density in the jam was approximately 0.78 bicycles/m$^2$. The density in the moving flow area was approximately 0.40 bicycles/m$^2$, and the flow rate was 0.55 bicycles/s/m. The wave propagation is 1.45 m/s, which is close to the 1.5 m/s observed in an experiment done by *(50)*.

## 5.3 Macroscopic sensitivity analysis: collision weight impact on capacity and maximum density for pedestrians and bicycles

Given the realistic trajectory patterns generated, a macroscopic analysis of the flow and density values as a function of different parameters may be conducted. In particular, the maximum flow and maximum density that can be achieved by the presented modeling framework are dependent on the collision weight parameter Wc. To study such dependency, we simulate bottleneck scenarios both for pedestrian and bicycles (separately) with focus on pedestrian to pedestrian and bicycle to bicycle interactions (collisions). Table 19 shows the results of the bicycle simulation and table 20 shows the results of the pedestrian simulation.

Table 19: Impact on Wc in the Bicycle Model.

| Scenario | Wc bike | Max Flow (B/m.s) | Max Density (b/m^2) |
|---|---|---|---|
| 1 | 1500 | 0.97 | 0.93 |
| 2 | 1600 | 0.91 | 0.81 |
| 3 | 1700 | 0.85 | 0.79 |
| 4 | 1800 | 0.79 | 0.74 |
| 5 | 1900 | 0.71 | 0.69 |
| 6 | 2000 | 0.63 | 0.63 |
| 6 | 2100 | 0.57 | 0.55 |
| 7 | 2200 | 0.55 | 0.49 |
| 8 | 2300 | 0.49 | 0.46 |
| 9 | 2400 | 0.42 | 0.41 |
| 10 | 2500 | 0.35 | 0.38 |

Table 20: Impact on Wc in the Pedestrian Model.

| Scenario | Wc ped | Max Flow (Ped/m.s) | Max Density (Ped/m^2) |
|---|---|---|---|
| 1 | 150 | 1.93 | 1.69 |
| 2 | 160 | 1.85 | 1.34 |
| 3 | 170 | 1.66 | 1.13 |
| 4 | 180 | 1.54 | 0.96 |
| 5 | 190 | 1.37 | 0.85 |
| 6 | 200 | 1.21 | 0.79 |
| 6 | 210 | 1.15 | 0.71 |
| 7 | 220 | 0.98 | 0.64 |
| 8 | 230 | 0.82 | 0.58 |
| 9 | 240 | 0.73 | 0.51 |
| 10 | 250 | 0.61 | 0.48 |

With the increasing values of the collision weighting parameter (Wc) value, the maximum flow, the maximum density and the maximum speed decrease. In the bike model, with the calibrated parameters from dataset 1 (associated with a mixed environment), we can reach a maximum density of 0.93 bike/$m^2$ and a maximum flow of 0.97 bike/m-sec. In the pedestrian model, with the calibrated parameters from dataset 1, we can reach maximum density of 1.69 ped/$m^2$ and a maximum flow of 1.93 ped/m-sec. These values should be taken with caution as the collision weights calibrated – even if associated with pedestrian to pedestrian or bike-to-bike interactions only – may be influenced by the specific scenario (set-up and traffic composition) relative perception of risks associated with different modes. For example, bikers may be over-estimating the weight of collisions when colliding with other bikers if traveling in a mixed setting - if compared to their weight of collisions if traveling only when bikers are present. With this same logic, pedestrians may be under-estimating the risk of collisions with other pedestrians if traveling in a mixed setting as they may be pre-occupied with the possibility of colliding with other cyclists. In summary, the impact of the collision weights on the maximum density/flow follow our expectation and the chosen Wc values produce realistic macroscopic measures' values. In addition to this first finding, the shape of the macroscopic traffic flow diagram with different collision weight values is clearly in line with previous work on particle dynamics - as may be seen in figure 21. Figure 21 shows the fitted curves rather than the entire scattered simulated flow-density points for better illustration. With the increase in Wc, the values of the maximum flow and maximum density clearly decreases while the change/reduction in speed is less obvious for pedestrians.

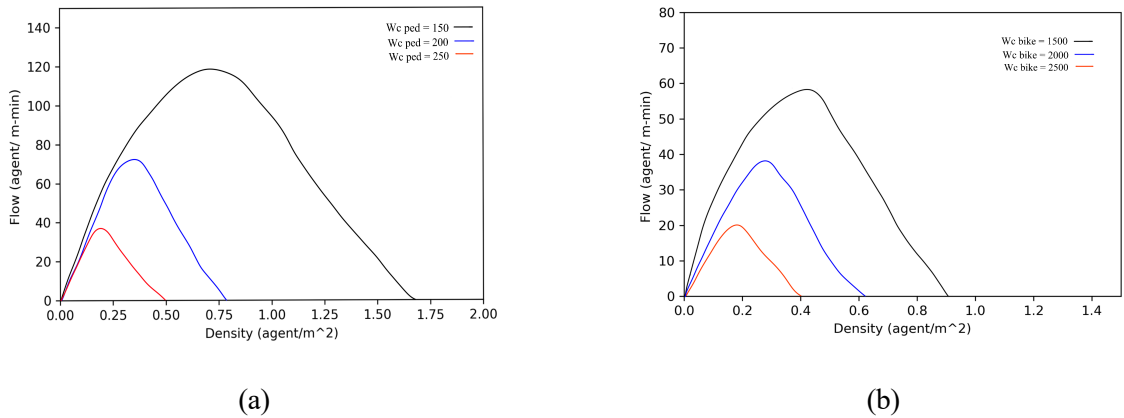

(a) (b)

Figure 21: Flow density relationship at different Wc values for Pedestrians (a) and bicyclists (b).

## 5.4 Simulated non-fitted fundamental diagrams in mixed simulation

A key objective of this study is to offer a single micro-economics based modeling framework to study the impact of the composition of micro-mobility modes on the traffic dynamics in a shared space environment. Once the different trajectory patterns and the main macroscopic measures are generated in a realistic manner, a more elaborate simulation analysis can be performed with a mix of pedestrians and cyclists and with different compositions. Accordingly, we set up a 5-minute simulation scenario at a bottleneck with the calibrated parameters values from Dataset 1 and with different percent of bicycles in the simulation. Figure 22 shows the flow density relationship for the different bicycle penetration rates.

From the flow density relationships of figure 22, it can be observed that - at low bike penetration rate - the mixed flow dynamics are governed by the pedestrians which tend to get closer to each other at lower speeds and thus allowing higher jam densities and higher capacities. With the increase of the bicycle penetration rate, the mixed flow dynamics start to be governed by slower bicyclists with a lesser ability to maneuver and larger size/occupied space by a bicycle. With this interesting outcome, at more than 40% bicycle penetration rate, the flow of pedestrians become heavily impacted by the bicyclists. A possible behavioral reason for such phenomena might be the cautious movement of pedestrians among bicyclists: pedestrians may need to be more aware of their surroundings and adjust their movements to accommodate the larger and faster-moving bicycles which leads to a lower flow rate for pedestrians. This kind of decrease in capacity continues when dealing with an-all bicycle traffic. In summary, there are very interesting unexplored dynamics when dealing with shared right of way among micro-mobility traffic modes of transportation. Such dynamics are dictated by individual perceptions and risk-taking tendencies with significant inter mode and inter user heterogeneity. Ultimately, some modes govern other modes depending on the type of trips made (e.g., commute versus recreational trips) and the type of scenario studied (e.g., bottleneck versus crossing scenarios). The modeling framework offered in this paper allows studying the generated traffic dynamics in such shared setting while accounting for the stochasticity and the cognitive dimensions in the human decision-making process.

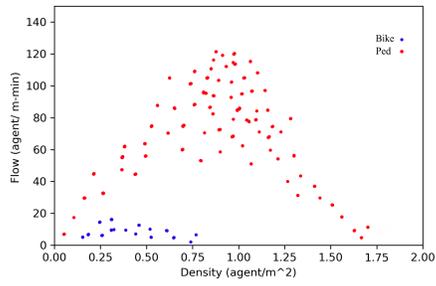

(a) Bike = 10%

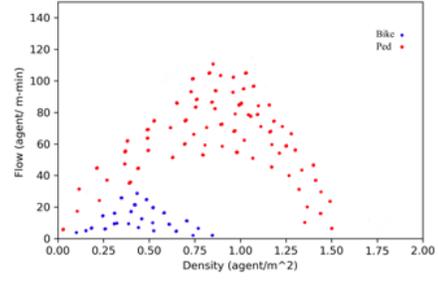

(b) Bike = 20%

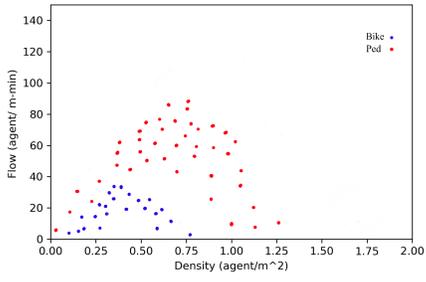

(c) Bike = 30%

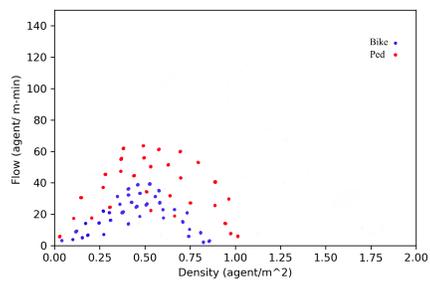

(d) Bike = 40%

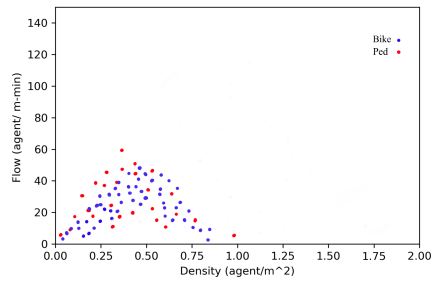

(e) Bike = 50%

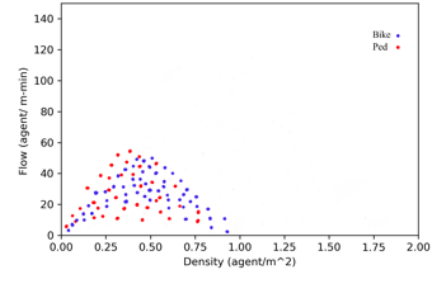

(f) Bike = 60%

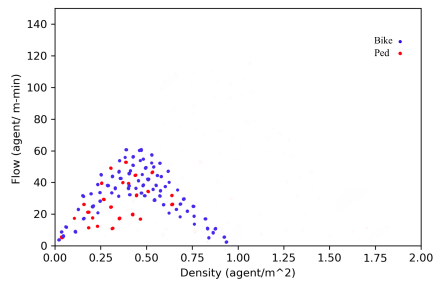

(g) Bike = 70%

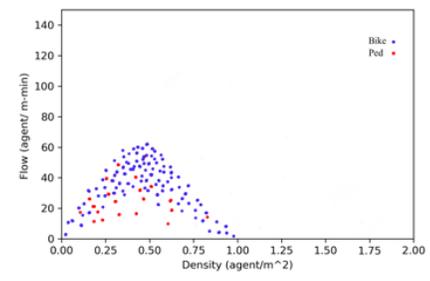

(h) Bike = 80%

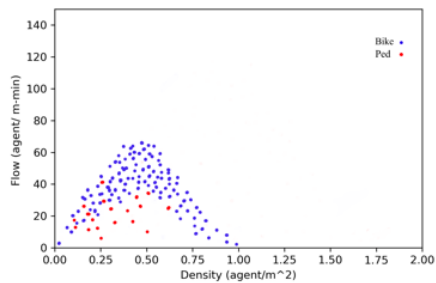

(i) Bike = 90%

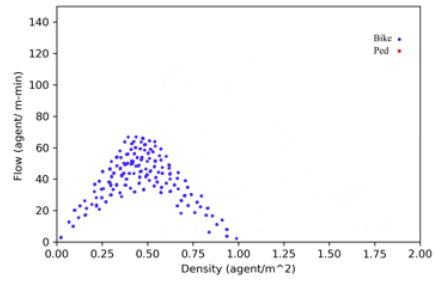

(j) Bike = 100%

Figure 22: Simulated flow density relationships at a bottleneck at different bike penetration rates.

In summary, we have compared our model with one of the latest mixed models for a pedestrian and bicycle and found that our model has a better performance in terms of RMSE and RRMSE values. The proposed model is also capable of generating some expected flow dynamics. The shape of the mixed bicycle and pedestrian flow fundamental diagrams at different bicycle penetration rate is also presented in this section.

# 6. Conclusion

A micro-economics based micro-mobility two-dimensional operational traffic model is formulated in this paper. The pedestrian and bicycle movement generated by this model is a balance between the desire of gaining speed reflected by a prospect theory value function and the consideration of colliding with other objects/road users reflected by a collision probability and its collision weight/seriousness. With such formation, this paper offers three key contributions. The first major contribution of this research is capturing pedestrian/cyclist behavior while incorporating roadway users' risk-taking attitude in the model equations. A parametric sensitivity analysis is presented to demonstrate the characteristics of the offered modeling paradigm especially when looking at the collision weight parameter and the stochastic nature of the velocities produced: the velocity distributions generated suggests that both pedestrians and bicyclists are risk averse in nature.

The second contribution of this research is data extraction and the calibration of the proposed model's parameters. We utilize an existing dataset from the Netherlands (Dataset 2) and we collect a new set of trajectories with a new dataset in the USA (Dataset 1). The two datasets are associated with two different naturalistic settings. To assess the properties of the proposed model, we perform the calibration exercise, and we use the calibrated parametric values to test the reproducibility of different behaviors and trajectory patterns. This study adopts the GA approach to calibrate the model while considering inter-agent heterogeneity. The calibration exercise shows the reflection of the models' parameters of the risk-taking tendencies of pedestrians and cyclists towards each other's while dealing with different contact/collision types. The values of these parameters may be seen as a new family of endogenous surrogate safety measures that are inherent to the proposed model's structure.

The third contribution of this study is the investigation of the macroscopic relationships generated by the individual stochastic interactions of different types of micro-mobility roadway users focusing on pedestrians and cyclists. The model was able to reproduce macroscopic relationships (in terms of shape) and fundamental measures (capacity, jam density …etc.) that are consistent with previous studies. At this stage of this research, simulation experiments can be conducted to recreate bottleneck dynamics, lane formation and evacuation phenomena. These scenarios can be expanded further with more in-depth safety and mobility analysis.

Some of the key findings of this study are as follows:

a) Pedestrian and the bicyclist weigh collision risks differently. Bicyclists are more sensitive (10-fold more sensitive) towards colliding with bicycles if compared to the sensitivity of pedestrians towards colliding with other pedestrians. Bicyclists' collision weight towards other bicyclists is more than the weight towards other pedestrians while pedestrians' collision weight towards other bicyclists is more that weight towards other pedestrians.

b) Both pedestrian and bicyclist are risk averse in nature. This is exhibited by the skewness towards lower velocities when studying the total utility distribution function associated with the velocity term when in contact with other users.

c) With the increase of the collision weight parameter, the subjects generally choose a lower speed but that does not necessarily lead to lower capacities. Accordingly, safer does not necessarily mean more efficient traffic system.

d) The proposed model can create pressure points when simulating pedestrian at very high densities such as

in case of evacuation through a narrow exit.

e) In shared spaces and in bottleneck conditions, a 40% or more bicycles' penetration rates result in traffic dynamics that are dominated by the bicyclists' characteristics despite the fact that there can be up to 60% of pedestrians in the mix. Below a 10 percent bicycles' penetration rate, the dynamics are dominated by pedestrians.

Despite the unique properties of the offered formulation and the associated promising results, some limitations may be addressed for further improvements and expansions. For example, the current formulation assumes that the collision probability is the maximum of all collision probabilities with all other agents and objects in the environment. A joint probability distribution might be suitable to explore but at the expense of the computational efficiency of the model. In addition, the collision seriousness term is assumed to be equal to 1. This might be changed by making the model sensitive to the collision characteristics with different objects. For example, the seriousness term may be dependent on the differential collision velocity term or the collision intensity. Finally, the model did show some lane formation phenomena. However, these lanes might need to be further studied as a function of the originating and existing points in a two-directional flow scenario.

In addition to addressing the possible limitations mentioned in the earlier paragraph, future studies, might focus on datasets with actual collision formation including collisions/accidents between pedestrians and bicycles and between pedestrians/bicycles and other motorized vehicles. With the proper dataset, the bicycle model presented in section 3 can be explored for e-scooterists' behavior modeling. The current model offers the possibility of simulating such scenarios with the possibility of including scooters. The relationship between collision related measures (such as time to collision and collision rates) and the risk-taking cognitive based model parameters may be explored further. Finally, the model can be readily extended to simulate cars and interaction of drivers with other micro-mobility modes in mixed urban settings.